\documentclass[11pt,twoside,a4paper,cmspaper,final,collab]{cms-tdr}
\def\svnVersion{a6dcb29}\def\svnDate{2026/03/12}\def\cmsCernNoTag{CERN-EP-2025-219}\def\cmsCernDate{\today}\def\cmsMessage{Published in Physical Review D as \href{http://dx.doi.org/10.1103/f9r8-styj}{\doi{10.1103/f9r8-styj}.}}
\begin{document}\cmsNoteHeader{SUS-23-016}

\newlength\cmsTabSkip\setlength{\cmsTabSkip}{1.0ex}
\ifthenelse{\boolean{cms@external}}
{\providecommand{\cmsTable}[1]{\relax{#1}}}
{\providecommand{\cmsTable}[1]{\resizebox{\textwidth}{!}{#1}}}
\newcommand{\currentL}{137}
\newcommand{\mpl}{\ensuremath{M_{\text{Pl}}}\xspace}
\newcommand{\mD}{\ensuremath{M_\text{D}}\xspace}
\newcommand{\mdm}{\ensuremath{M_{\text{DM}}}\xspace}
\newcommand{\mmed}{\ensuremath{M_{\text{med}}}\xspace}
\newcommand{\gDM}{\ensuremath{g_\text{DM}}\xspace}
\newcommand{\gl}{\ensuremath{g_{l}}\xspace}
\newcommand{\gq}{\ensuremath{g_{\PQq}}\xspace}
\newcommand{\ptmissvec}{\ensuremath{\vec{p}_T^{\text{miss}}}\xspace}
\newcommand{\dR}{\ensuremath{\Delta R}\xspace}
\newcommand{\zinvg}{\ensuremath{\PZ({\to}\PGn\PAGn){+}\PGg}\xspace}
\newcommand{\zllg}{\ensuremath{\PZ({\to}\Pell\PAell){+}\PGg}\xspace}
\newcommand{\wlng}{\ensuremath{\PW({\to}\Pell\PAGn){+}\PGg}\xspace}
\newcommand{\gj}{\ensuremath{\PGg{+}\text{jets}}\xspace}
\newcommand{\ttg}{\ensuremath{\ttbar\PGg}\xspace}
\newcommand{\tg}{\ensuremath{\PQt\PGg}\xspace}
\newcommand{\ETg}{\ensuremath{E_{\mathrm{T}}^{\PGg}}\xspace}
\newcommand{\pTgvec}{\ensuremath{\vec{p}_{\mathrm{T}}^{\PGg}}\xspace}
\newcommand{\sigetaeta}{\ensuremath{\sigma_{i\eta i\eta}}}
\newcommand{\DphiMETgvec}{\ensuremath{\Delta\phi(\ptmissvec, \pTgvec)}\xspace}
\newcommand{\minDphiMETj}{\ensuremath{\text{min}\Delta \phi(\ptvecmiss, \ptvec^{\mkern3mu\text{jet}})}\xspace}
\newcommand{\NNPDFthreeone}{NNPDF3.1\xspace}
\newcommand{\DYRes}{\textsc{DYRes}\xspace}
\newcommand{\PGgst}{{\HepParticle{\PGg}{}{\ast}}\Xspace}
\providecommand{\PAGc}{{\HepAntiParticle{\PGc}{}{}}\xspace}

\cmsNoteHeader{SUS-23-016}

\title{Search for new physics in the final state with a single photon and large missing transverse momentum in proton-proton collisions at \texorpdfstring{$\sqrt{s} = 13\TeV$}{sqrt(s) = 13 TeV}}

\author[cern]{The CMS Collaboration}
\date{\today}

\abstract{
A search for new physics in events featuring a single photon and missing transverse momentum is presented, using proton-proton $\sqrt{s} = 13~\TeV$ collision data corresponding to an integrated luminosity of 101\fbinv collected by the CMS experiment at the CERN LHC between 2017 and 2018. This analysis, combined with a previous study of 36\fbinv of 2016 data (totaling 137\fbinv), reveals no significant deviations from standard model expectations. The results are then used to establish 95\% confidence level limits on parameters in theoretical models involving dark matter and large extra dimensions. Compared to the 2016-only analysis, this search achieves up to a 14\% improvement in exclusion reach for mediator masses in simplified dark matter models, along with 11\% and 10\% enhancements in the limits on the effective field theory suppression scale and the fundamental Planck scale, respectively. These results are the most stringent constraints on these parameters to date.}

\hypersetup{
pdfauthor={CMS Collaboration},
pdftitle={Search for new physics in the final state with a single photon and large missing transverse momentum in proton-proton collisions at sqrt s = 13 TeV},
pdfsubject={CMS},
pdfkeywords={CMS, dark matter, gravitons}}

\maketitle

\section{Introduction}
\label{sec:introduction}

This paper details a search by the CMS experiment at the CERN LHC for a distinctive signature of new physics: events with a high-energy photon and a substantial imbalance in transverse momentum. This ``monophoton'' final state serves as a crucial probe for various theoretical scenarios, including those that feature large extra dimensions or contain dark matter (DM).

\begin{figure*}[hbtp]
  \centering
    \includegraphics[width=0.32\textwidth]{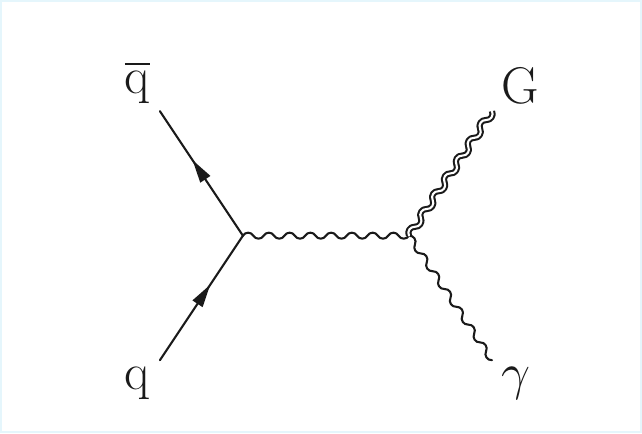}
    \includegraphics[width=0.32\textwidth]{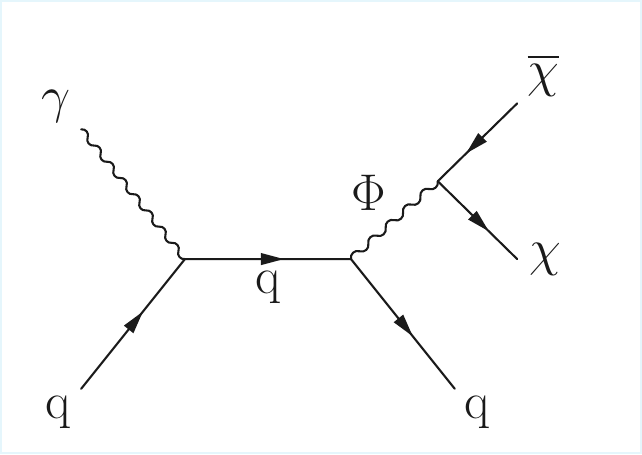}
    \includegraphics[width=0.32\textwidth]{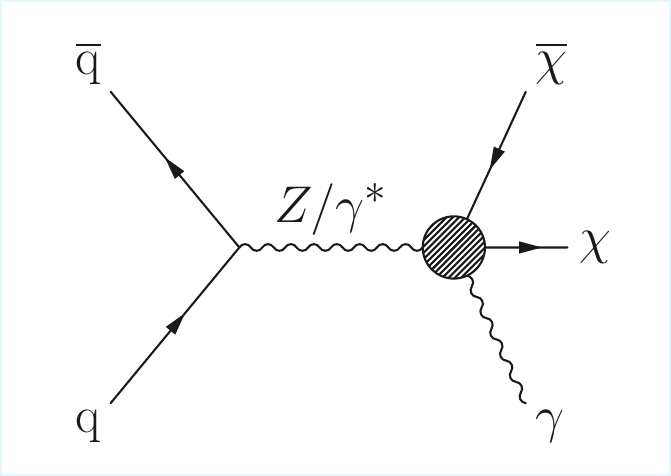}

    \caption{
      Leading-order diagrams of graviton production in the ADD model (left), simplified DM model (center), and EW--DM effective interaction (right), with a final state comprising a photon and large \ptmiss. Particles \PGc and \PAGc are the DM and its antiparticle, and $\Phi$ in the simplified DM model represents a vector or axial-vector mediator.
    }
    \label{fig:diagrams}
\end{figure*}

The Arkani-Hamed, Dimopoulos, and Dvali (ADD) model~\cite{ADD,ADD1,Arkani-Hamed:1999wga} postulates the existence of a number of large, compactified extra spatial dimensions ($n$). This framework addresses the hierarchy problem by introducing a lower fundamental Planck scale (\mD) in the higher dimensional space, while the effective Planck scale (\mpl) observed in our three-dimensional space is much larger due to propagation of gravity into the extra dimension. This lowering of the Planck scale allows for the potential production of Kaluza-Klein (KK) gravitons. The ADD model predicts the production of a photon and one or more non-interacting KK gravitons(\cPG) via $\Pq\Paq\to \PGg\cPG$, as shown in Fig.~\ref{fig:diagrams} (left), where each graviton has a mass up to \mD. The undetected gravitons result in a significant missing transverse momentum (\ptmiss) in the detector.

At the LHC, DM particles are expected to be produced in high-energy collisions and detected indirectly through their interactions with standard model (SM) particles. Since DM is expected to be weakly interacting and invisible to the detectors, these searches typically look for \ptmiss in association with SM particles~\cite{Beltran:2010ww,Goodman:2010ku,Fox:2011pm}. Simplified models recommended by Ref.~\cite{Abercrombie:2015wmb} are employed to interpret these signatures. These models provide extensions to the SM by introducing a DM particle and a heavy mediator, as shown in Fig.~\ref{fig:diagrams} (center). The mediator can be either a vector or an axial vector boson. It couples to the incoming quark through a coupling constant \gq = 0.25, and to DM particles with coupling constant \gDM = 1.0, with coupling to leptons, \gl, fixed to zero~\cite{Boveia:2016mrp}. These coupling values are chosen
to ensure that the mediator remains within the narrow-width approximation, where the mediator width is small compared to its mass. This approximation allows the mediator to be treated as a short-lived particle with a well-defined mass, ensuring that the signal modeling remains valid and enabling consistent comparisons across different experimental searches.

The DM particles can also be produced through effective operators involving photons. This interaction is known as the electroweak dark matter (EW-DM) contact interaction and is characterised by the process $\PQq\PAQq\to\PZ/\PGgst\to\PGg\PGc\PAGc$~\cite{Nelson:2013pqa}, where \PGc is the DM particle, as shown in Fig.~\ref{fig:diagrams} (right). This interaction is parametrized by two dimensionless couplings, $k_{1}$ and $k_{2}$, which control the relative strength of the DM interaction with the EW gauge bosons. The constants $k_{1}$ and $k_{2}$ are the couplings to the $U(1)_Y$ and $SU(2)_L$ gauge bosons, respectively. The overall interaction is suppressed by a parameter $\Lambda$, such that the effective coupling scales as $k_{1}/\Lambda$ and $k_{2}/\Lambda$. The parameters defined in these models are the same as those described in a previous CMS search analyzing 2016 data~\cite{Sirunyan:2017ewk}.

The analysis targets events with a single high-energy photon and significant \ptmiss, a characteristic signature of several new physics scenarios. The dominant SM backgrounds include \zinvg, \wlng, and \gj processes, while other minor background contributions are \zllg, \PGg{}\PGg, \tg, \ttg, $\PV\PV\PGg$ (where \PV refers to a \PW or \PZ boson), and beam halo events.

A similar search in $\Pp\Pp$ collisions at $\sqrt{s} = 13\TeV$, based on a data set corresponding to an integrated luminosity of 139\fbinv, has been reported by the ATLAS experiment~\cite{ATLAS:2020uiq}. No significant excess over the SM prediction was observed. In that search, the observed lower limit on the axial-vector (vector) mediator mass was determined to be 1460 (1470)\GeV for low DM masses, assuming the same \gq, \gDM, and \gl coupling values.

While the ATLAS search focuses solely on simplified models for DM, the CMS search also explores the ADD model of large extra dimensions and the EW-DM framework. The previous search in the same final state by the CMS experiment~\cite{Sirunyan:2017ewk} was based on data collected during 2016 corresponding to an integrated luminosity of 36\fbinv. In this paper, we analyze 2017 and 2018 data, and combine with the previously published result using 2016 data~\cite{Sirunyan:2017ewk}. The combined data set corresponds to a total integrated luminosity of \currentL \fbinv. The results presented here supersede those of the previous search.

The structure of this paper is as follows. Section~\ref{sec:detector} provides a description of the CMS detector apparatus, along with the algorithm employed to reconstruct particles in pp collision events within the detector. Section~\ref{sec:event_selection} outlines the criteria that events must meet to be included in the signal region (SR) and control region (CR). Section~\ref{sec:modeling} lists the generators used to model signal and background processes. Section~\ref{sec:bkgestimate} describes the methods used to estimate the expected background yields in the SR and CR. Systematic uncertainties are discussed in Section~\ref{sec:systematic}. The results and interpretations are presented in Section~\ref{sec:results}, and the overall conclusions are summarized in Section~\ref{sec:summary}. The tabulated results are also provided in the HEPData record~\cite{hepdata}.

\section{The CMS detector and event reconstruction}
\label{sec:detector}

The CMS apparatus~\cite{CMS:2008xjf,CMS:2023gfb} is a multipurpose, nearly hermetic detector, designed to trigger on and identify electrons, muons, photons, and charged and neutral hadrons~\cite{CMS:2020uim,CMS:2018rym,CMS:2014pgm}.  Its central feature is a superconducting solenoid of 6\unit{m} internal diameter, providing a magnetic field of 3.8\unit{T}. Within the solenoid volume are a silicon pixel and strip tracker, a lead tungstate crystal electromagnetic calorimeter (ECAL), and a brass and scintillator hadron calorimeter (HCAL), each composed of a barrel and two endcap sections. Forward calorimeters extend the pseudorapidity coverage provided by the barrel and endcap detectors. Muons within the pseudorapidity range $\abs{\eta} < 2.4$ are reconstructed using gas-ionization detectors embedded in the steel flux-return yoke outside the solenoid. More detailed descriptions of the CMS detector, together with a definition of the coordinate system used and the relevant kinematic variables, can be found in Refs.~\cite{CMS:2008xjf,CMS:2023gfb}.

Events of interest are selected in real time using a two-tiered trigger system. The first level, composed of custom hardware processors, uses information from the calorimeters and muon detectors to select events at a rate of around 100\unit{kHz} within a fixed latency of about 4\mus ~\cite{CMS:2020cmk}. The second level, known as the high-level trigger, consists of a farm of processors running a version of the full event reconstruction software optimized for fast processing, and reduces the event rate to approximately 1\unit{kHz} before data storage~\cite{CMS:2016ngn,CMS:2024aqx}.

The primary vertex is taken to be the vertex corresponding to the hardest scattering in an event, evaluated using tracking information alone, as described in Section 9.4.1 of Ref.~\cite{Contardo:2020886}.
The global event reconstruction (also called particle-flow (PF) event reconstruction~\cite{CMS:2017yfk}) aims to reconstruct and identify each individual particle in an event, with an optimized combination of all subdetector information. In this process, the identification of the particle type (photon, electron, muon, charged hadron, neutral hadron) plays an important role in the determination of the particle direction and energy. Photons are identified as ECAL energy clusters not linked to the extrapolation of any charged-particle trajectory to the ECAL. Electrons are identified as a primary charged-particle track and potentially many ECAL energy clusters corresponding to the extrapolation of this track to the ECAL and to possible bremsstrahlung photons emitted along the way through the tracker material. Muons are identified as tracks in the central tracker consistent with either a track or several hits in the muon system, and associated with calorimeter deposits compatible with the muon hypothesis. Charged hadrons are identified as charged-particle tracks that are identified neither as electrons nor as muons. Finally, neutral hadrons are identified as HCAL energy clusters not linked to any charged-hadron trajectory, or as a combined ECAL and HCAL energy excess with respect to the expected charged-hadron energy deposit. 

The energy of photons is obtained from the ECAL measurement. The energy of electrons is determined from a combination of the track momentum at the main interaction vertex, the associated ECAL cluster energy, and the energy sum of all bremsstrahlung photons attached to the track. Muon energy is determined from the track momentum. The energy of charged hadrons is determined from a combination of the track momentum and the corresponding ECAL and HCAL energies, corrected for the response function of the calorimeters to hadronic showers. Finally, the energy of neutral hadrons is obtained from the corrected ECAL and HCAL energies associated with those particles.
For each event, hadronic jets are clustered from these reconstructed particles using the infrared and collinear safe anti-\kt algorithm ~\cite{Cacciari:2008gp, Cacciari:2011ma} with a distance parameter of 0.4. Jet momentum is determined as the vectorial sum of all particle momenta in the jet, and is found from simulation to be, on average, within 5 to 10\% of the true momentum over the entire \pt spectrum and detector acceptance. Additional pp interactions within the same or nearby bunch crossings (pileup) can contribute additional tracks and calorimetric energy depositions to the jet momentum. To mitigate this effect, charged-particles identified as originating from pileup vertices are discarded, and an offset correction is applied to correct for remaining contributions~\cite{CMS:2016lmd}.

Jet energy corrections are derived from simulation to bring the measured detector response of jets to that of particle-level jets on average. In situ measurements of the momentum balance in dijet, $\PGg + \text{jet}$, $\PZ + \text{jet}$, and multijet events are used to account for any residual differences in the jet energy scale between data and simulation ~\cite{CMS:2016lmd}. The jet energy resolution amounts typically to 15--20\% at 30\GeV, 10\% at 100\GeV, and 5\% at 1\TeV ~\cite{CMS:2016lmd}. Additional selection criteria are applied to each jet to suppress jets with anomalous energy deposits, noise, or reconstruction failures in specific subdetector components, such as the calorimeters or tracker~\cite{CMS:2019ctu,CMS:2017wyc}.

The missing transverse momentum vector \ptvecmiss is computed as the negative vector sum of the transverse momenta of all the PF candidates in an event, and its magnitude is denoted as \ptmiss ~\cite{CMS:2019ctu}. The \ptvecmiss is modified to account for corrections to the energy scale of the reconstructed jets in the event. Anomalous high-\ptmiss events can arise from a variety of reconstruction failures, detector malfunctions, or noncollision backgrounds. Such events are rejected by event filters that are designed to identify more than 85--90\% of the spurious high-\ptmiss events with a mistagging rate of less than 0.1\% ~\cite{CMS:2019ctu}.

In the barrel section of the ECAL, an energy resolution of approximately 1\% is achieved for unconverted or late-converting photons at energies of $\mathcal{O}(10\,\mathrm{\GeV})$. The energy resolution of the remaining barrel photons is about 1.3\% up to $\abs{\eta} = 1$, changing to about 2.5\% at $\abs{\eta} = 1.4$ ~\cite{CMS:2015myp}.

\section{Event selection}
\label{sec:event_selection}

The integrated luminosity of the analyzed data sample is 101.3 \fbinv. The data sample was collected using a single-photon trigger that required at least one photon candidate with $\pt> 200\GeV$. The trigger efficiency is measured to be about 95\% for events passing the analysis selection.
Events are selected from the recorded data by requiring $\ptmiss> 200\GeV$ and the presence of at least one photon with $\ETg >225\GeV$ (where \ETg is the transverse energy of the photon) within the fiducial region of the ECAL barrel ($\abs{\eta} < 1.44$). The analysis is restricted to the barrel region due to its superior energy resolution and lower background contamination compared to the endcaps, as well as the limited expected signal contribution in the endcap region. The final sample contains no events with more than one photon meeting the selection criteria.

Photon candidates are identified using calorimetric information, isolation, and the absence of an electron seed. The first two criteria help distinguish genuine photon candidates from electromagnetic (EM) showers originating from hadrons, while the third criterion ensures the photon candidates are not misidentified electrons. The calorimetric requirements for photons consist of the ratio of energy deposited in the HCAL to that in the ECAL ($H/E < 0.023$) and $\sigetaeta < 0.0103$. The variable \sigetaeta, described in detail in Ref.~\cite{CMS:2015myp}, represents the width of the EM shower in the $\eta$ direction, which is generally larger in showers from hadronic activity. For a photon candidate to be considered as isolated, the scalar sums of the transverse momenta of charged hadrons, neutral hadrons, and photons within a cone of $\dR = \sqrt{\smash[b]{(\Delta \eta)^2 + (\Delta \phi)^2}} < 0.3$ around the candidate photon must fall below a set of corresponding bounds chosen to give 80\% signal efficiency. Only the PF candidates that do not overlap with the EM shower of the candidate photon are included in the isolation sums. The charged hadron isolation ($I_{\text{ch}}$) is defined as the scalar sum of the \pt of charged hadrons within the cone. The neutral hadron isolation ($I_{\text{nh}}$) is the scalar sum of the \pt of neutral hadrons within the cone, and the photon isolation ($I_{\text{ph}}$) is the sum of \pt of photons within the cone, excluding the photon itself. Ideally, the isolation sum over PF charged hadrons should be computed using only the candidates sharing an interaction vertex with the photon candidate. However, because photon candidates are not reconstructed from tracks, their vertex association is undefined. When an incorrect vertex is assigned, nonisolated photon candidates can appear isolated. To reduce the rate for accepting nonisolated photon candidates, the maximum charged-hadron isolation ($I_{\text{ch}}^{\text{max}}$) value over all vertex hypotheses (worst isolation) is used. The above criteria select efficiently both unconverted photons and photons undergoing conversion in the detector material in front of the ECAL.

Stray ECAL clusters produced by mechanisms other than $\Pp\Pp$ collisions can be misidentified as photons. Beam halo muons in particular, which accompany the proton beams and traverse the detector longitudinally, contribute significantly to the rate of false photon candidates. 
To reject this background, the maximum of the total calorimeter energy summed along all possible paths of beam halo particles passing through the cluster (halo total energy), calculated for each photon candidate, must be below 4.9\GeV. To better constrain this background, the signal region is split into two parts based on the azimuthal angle $\phi$ of the photon. The region with $\abs{\phi} < 0.5$ is called the horizontal region, and the rest is called the vertical region. This helps in improving the estimation of the beam halo contribution, as discussed later in Section~\ref{subsec:halo}.

\begin{figure*}[hbtp]
    \centering
    \includegraphics[width=0.48\textwidth]{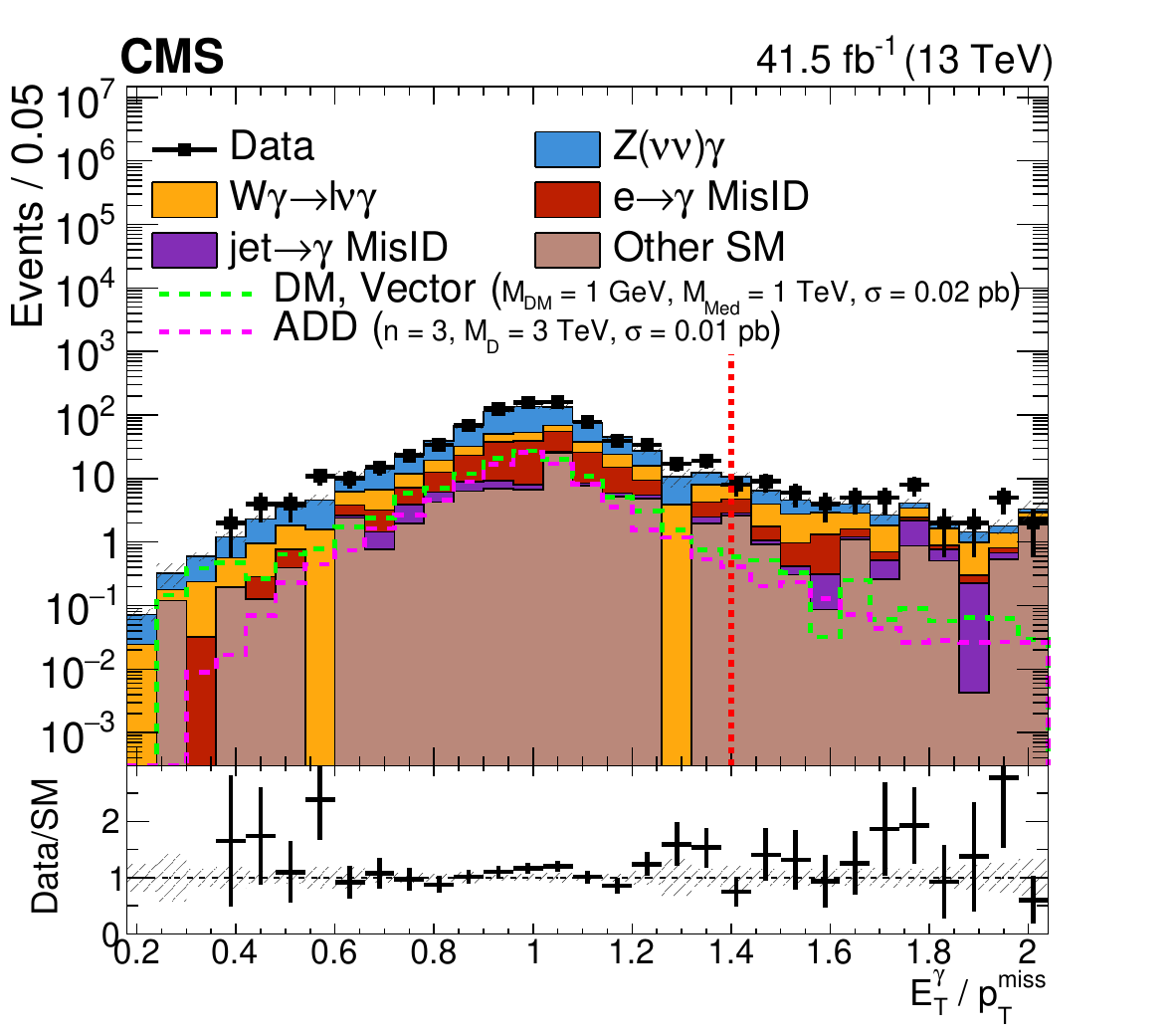}
    \includegraphics[width=0.48\textwidth]{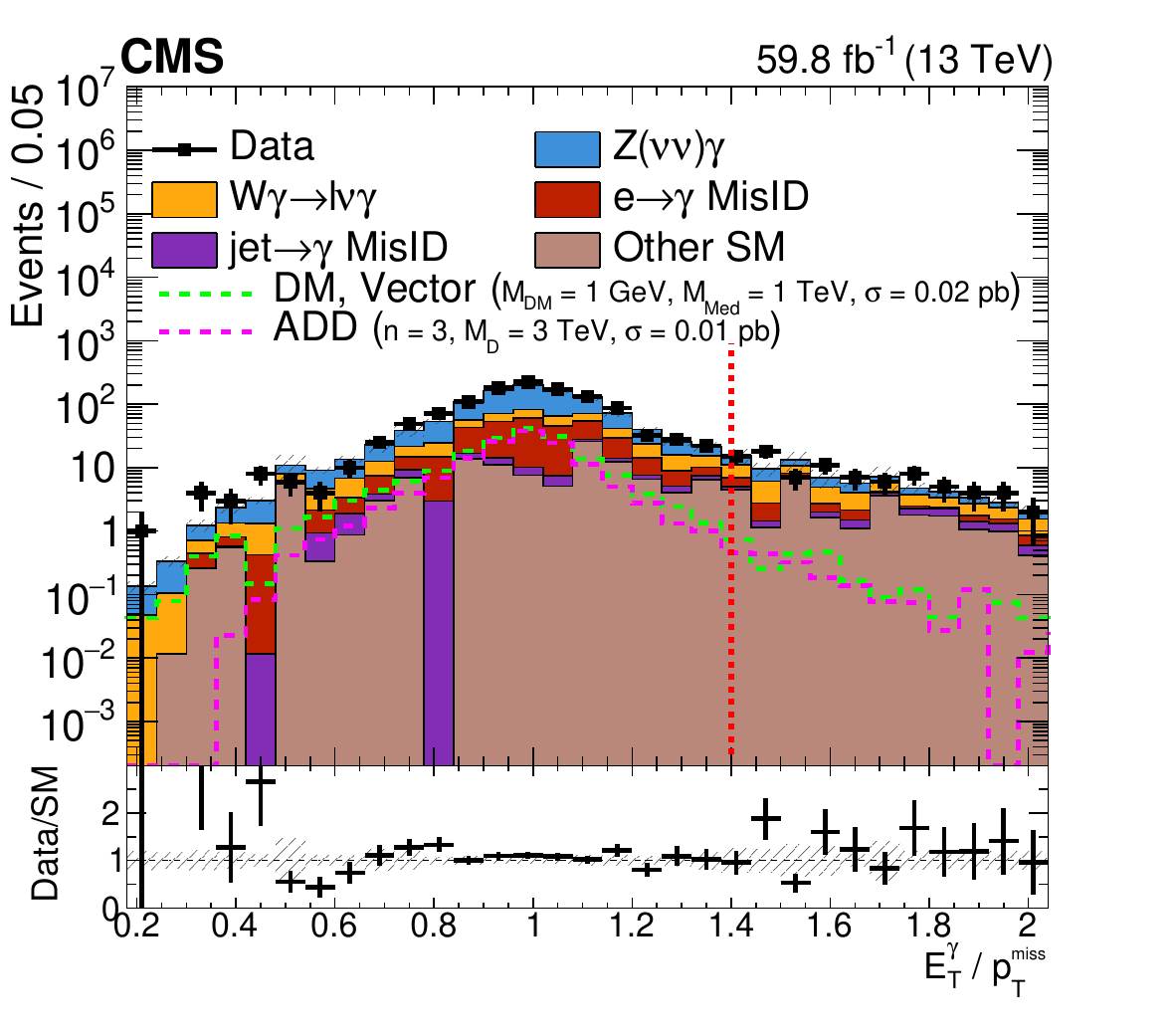}

    \caption{Distribution of \(\ETg/\ptmiss\) for the 2017 (left) and 2018 (right) data sets. Templates for signal hypotheses are shown overlaid as light green and magenta dashed lines along with their cross section values. The cross hatched band represents the total systematic and statistical uncertainties. Events to the right of the red dashed vertical line are excluded.}
    \label{fig:Pt_MET_prefit}
\end{figure*}

Events with a high-\pt photon and large \ptmiss are subjected to further requirements to suppress SM background processes that feature a genuine high-energy photon, but not a significant amount of \ptmiss. One such SM process is \gj, where an apparent large \ptmiss is often the result of a mismeasured jet energy.
In contrast to signal processes, \ptmiss is typically smaller than \ETg in these events, so requiring the ratio of \ETg to \ptmiss to be less than 1.4 effectively rejects 63\% of the \gj background while retaining 99\% of the signal efficiency. Fig.~\ref{fig:Pt_MET_prefit} shows the comparison of the observed and simulated \ETg /\ptmiss distribution.

Events are also rejected if the minimum opening angle between \ptvecmiss and the direction of any of the (up to four) highest-\pt jets, \minDphiMETj, is less than 0.5 radians. Only jets with $\PT > 30\GeV$ and $\abs{\eta} < 5$ are considered in the \minDphiMETj calculation. The expected and observed numbers of jets in the SR are small, typically zero or one. For the \gj process, rare and anomalous mismeasurement of \ETg can also lead to large \ptmiss. For this reason, the photon candidate transverse momentum vector (\pTgvec) and \ptvecmiss must be separated by more than 0.5 radians.

\begin{figure*}[hbtp]
    \centering
    \includegraphics[width=0.48\textwidth]{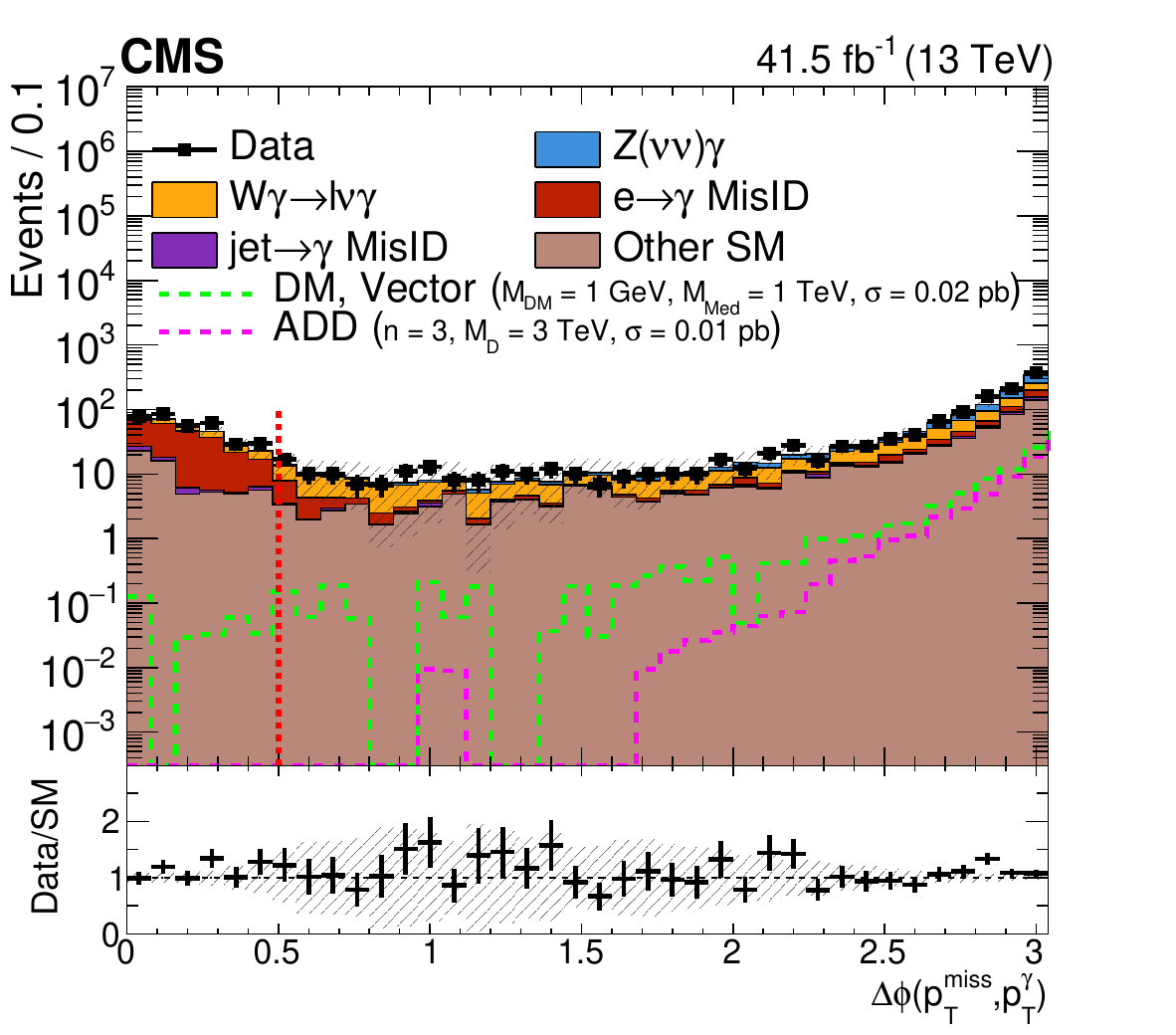}
    \includegraphics[width=0.48\textwidth]{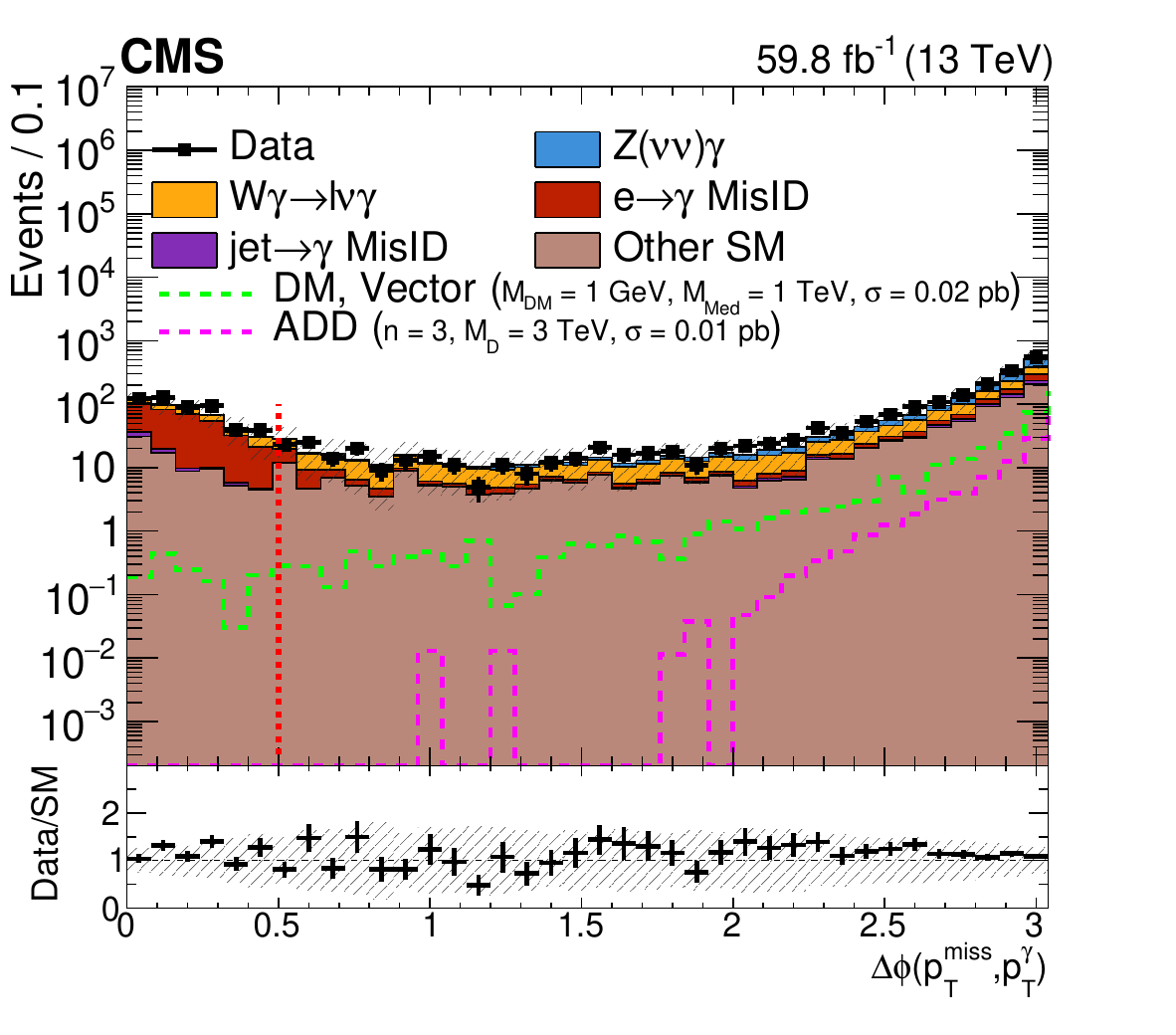}
        \caption{Distribution of \DphiMETgvec for the 2017 (left) and 2018 (right) data sets. Templates for signal hypotheses are shown overlaid as light green and magenta dashed lines along with their cross section values. The cross hatched band represents the total systematic and statistical uncertainties. Events to the left of the red dashed vertical line are excluded.}

        \label{fig:DeltaPhi_Pho_MET}
\end{figure*}

This is illustrated in Fig.~\ref{fig:DeltaPhi_Pho_MET}, which shows the \DphiMETgvec distribution before any selection on this variable is applied. This requirement effectively suppresses backgrounds from QCD multijet events, electron to photon misidentification and \gj events.

During the 2018 data-taking period, a section of the HCAL was not working, leading to irrecoverable mismeasurement in a localized region of the detector ($-1.57 < \phi <-0.87$, $-3.0<\eta<-1.3$). To avoid contamination from such mismeasurement, events containing any jet with $\PT > 30\GeV$ in this region are rejected.

The SM \wlng process is reduced by vetoing events in which an electron or a muon with $\PT > 10\GeV$  is separated from the photon by $\dR > 0.5$. Electrons and muons are selected using ``Loose'' identification (ID) criteria that include isolation and other quality requirements~\cite{CMS:2020uim,CMS:2019ied}.  The event selection criteria for the SR are summarized in Table~\ref{tab:event-selection1}.

\begin{table*}[!hbtp]
\centering
\topcaption{Event selection criteria for the SR. The requirements on the trigger, photon identification, and $\Delta\phi$ are listed.}
\label{tab:event-selection1}
\renewcommand{\arraystretch}{1.3}
\begin{scotch}{ll}
Category & SR \\[\cmsTabSkip]
\hline
Trigger                    & $\ETg > 200\GeV$ \\
\PGg requirement         & $\ETg > 225\GeV$, $\abs{\eta} < 1.44$\\
                           & passing 80\% signal efficiency selection: $\sigma_{i\eta i\eta} < 0.0103$,\\
                           & $H/E < 0.023$, $I_{\text{ch}}^{\text{max}} < 2.14\GeV$, \\
                           & $I_{\text{ph}} < 0.17\GeV + 0.003 \ETg$, \\
                           & $I_{\text{nh}} < 7.25\GeV + 0.01 \ETg + 0.00001 (\ETg)^2/\GeV$ \\
\ptmiss requirement            & $\ptmiss > 200\GeV$ \\
Angular separation         & $\DphiMETgvec > 0.5$ \\
Jet requirement            & $\Delta\phi(\ptvec^{\text{jet}}, \ptmissvec) > 0.5$ (for the leading 4 jets) \\
$\PGg$+Jets suppression  & $\ETg / \ptmiss < 1.4$ \\
Lepton veto & No \PGmm or \Pem with $p_T > 10\GeV$ and passing Loose ID criteria \\

\end{scotch}
\end{table*}

To estimate the dominant backgrounds and constrain their normalization in data, several dedicated CRs are defined. The single-lepton CRs (single-electron and single-muon) are designed to constrain the \wlng background, while the dilepton CRs (dielectron and dimuon) constrain the yields of \zllg events, which serve as a proxy for the \zinvg background.

The single-electron (single-muon) CR is defined by a requirement of exactly one tight electron (muon) with $\PT > 30\GeV$ and $\abs{\eta} < 2.5\,(2.4)$, in addition to a photon requirement that is identical to the one for the signal region (SR). To suppress the contributions from large-\ptmiss processes other than \wlng, the transverse mass $m_{T}$ = $\sqrt{\smash[b]{2\ptmiss\PT^{\Pell}[1-\cos\Delta\phi(\ptvecmiss,\ptvec^{\Pell})]}}$ must be less than $160\GeV$, where $\PT^{\Pell}$ denotes the transverse momentum of the selected lepton (electron or muon). Additionally, for the single-electron CR, \ptmiss must be greater than 50\GeV to limit the contribution from the \gj process, where a jet is misidentified as an electron. Finally, the recoil vector $\vec{U} = \ptvecmiss + \ptvec^{\Pell}$, which serves as this region's analogue for \ptvecmiss in the SR, must satisfy identical requirements to those for the \ptvecmiss in the SR.

The dielectron (dimuon) CR is defined by exactly two electrons (muons) of opposite electric charge, in addition to the photon, with $60 < m_{\Pell\Pell} < 120\GeV$, where $m_{\Pell\Pell}$ is the invariant mass of the dilepton system. The recoil vector of this region is $\vec{U} = \ptvecmiss + \ptvec^{\Pell\Pell}$, where the transverse momentum of the dilepton system is given by $\ptvec^{\Pell\Pell} = \ptvec^{\Pell_1} + \ptvec^{\Pell_2}$ and must satisfy identical requirements to those for the \ptvecmiss in the SR.
The event selection criteria for the CRs are summarized in Table~\ref{tab:event-selection2}.

\begin{table*}[!htb]
\centering
\topcaption{Event selection in the CRs.}
\label{tab:event-selection2}
\cmsTable{
\begin{scotch}{lll}
Category &
Single Lepton (\wlng) &
Double Lepton (\zllg) \\[\cmsTabSkip]
\hline
Trigger, Photon, $\Delta\phi(\text{jet}, \ptmiss)$ & Same as SR & Same as SR \\[0.5\cmsTabSkip]
Lepton & $\pt  > 30\GeV$,  & Leading: $\pt  > 30\GeV$,\\
&passing Tight ID requirement&passing Tight ID requirement\\
&&Subleading: $\pt  > 10\GeV$, \\
&&opposite charge,  \\
&&passing Loose ID requirement\\[0.5\cmsTabSkip]
Recoil ($U$) &
$\abs{U} > 200\GeV$,   &
$\abs{U} > 200\GeV$,  \\
& $\Delta\phi(\pTgvec, \vec{U}) > 0.5$ & $\Delta\phi(\pTgvec, \vec{U}) > 0.5$ \\[0.5\cmsTabSkip]
Additional requirements & 
$m_{\mathrm{T}}(\ptvec^{\Pell} \ptvecmiss) < 160\GeV$,  &
$60 < m_{\Pell\Pell} < 120\GeV$ \\
&$\ptmiss > 50\GeV$ (only for \PW{}(\Pe{}\PGn))&
\end{scotch}
}
\end{table*}

\section{Simulated samples}
\label{sec:modeling}

Simulated predictions for signal and some backgrounds derive from Monte Carlo (MC) samples, with the \NNPDFthreeone NNLO~\cite{NNPDF:2017mvq} parton distribution function (PDF) set and the strong coupling constant value set to $\alpha_{S} = 0.118$.

The simulated events for the \zinvg, \zllg, and \wlng SM processes are generated using the \MGvATNLO{} generator at LO, with up to two extra partons in the matrix element calculations~\cite{Alwall:2014hca}. These are then normalized to the next-to-LO electroweak (NLO EW) and next-to-NLO quantum chromodynamics (NNLO QCD) cross sections using the correction factors described in Section~\ref{sec:bkgestimate}.  

Parton showering and hadronization are modeled by \PYTHIA{} 8.240~\cite{Sjostrand:2014zea} with the underlying-event tune CP5~\cite{CMS:2015wcf}. Multiple simulated minimum bias events are overlaid on the primary interaction to model the distribution of pileup in data. Generated particles are processed through the full \GEANTfour-based simulation of the CMS detector~\cite{GEANT, GEANTdev}.

For the DM signal hypotheses, \MGvATNLO is used to produce MC simulated samples at NLO in QCD, requiring $\ETg > 130\GeV$ and $\abs{\eta^{\PGg}} <2.5 $. A large set of DM simplified model samples are generated, with mediator mass (\mmed) ranging from 1--2000\GeV and dark matter mass (\mdm) ranging from 1--1000\GeV. In the same way, EW-DM effective interaction samples are generated at LO in a range of 1--1000\GeV for the DM particle mass. The ADD signal samples are generated at LO using \PYTHIA{} 8.240, requiring $\ETg > 130\GeV$, with no restriction on the photon $\eta$. Samples are generated in a grid of values for the $n = 3\text{--}6$ and $\mD = 1\text{--}6~\TeV$ scenarios. The efficiency of the full event selection for these signal models ranges from 6 to 29\% for the simplified DM models, from 44 to 46\% for the EW-DM model, and from 23 to 30\% for the ADD model, depending on the parameters of the models.

\section{Background estimation}
The primary irreducible background for the $\PGg+\ptmiss$ signature is SM \PZ boson production in association with a photon, \zinvg. Other SM background processes include \wlng (where the charged lepton $\Pell$ is undetected), $\PW\to\Pell\Pgn$ (where $\Pell$ is falsely identified as a photon), \gj, QCD multijet events (with a jet misidentified as a photon), \PGg{}\PGg, t\PGg, \ttg, VV\PGg (where V refers to a \PW or a \PZ boson), and \zllg. In addition, a small residual number of events from noncollision sources, such as beam halo interactions~\cite{Beamhalo}, contribute to the total background. The background estimation and signal extraction are performed concurrently via a simultaneous maximum likelihood (ML) fit across the signal region and the corresponding single-lepton and dilepton CRs. A global likelihood function is constructed to model the expected background contributions in each bin of the \ETg variable in both SR and CRs, as well as the expected signal yield in each SR bin.  The best-fit background model and signal strength are obtained by maximizing the likelihood function. The likelihood function is defined in the same way as described in Ref.~\cite{Sirunyan:2017ewk}. Separate approaches are adopted to estimate the dominant \zinvg and \wlng backgrounds, and subdominant backgrounds, as detailed below. 

\label{sec:bkgestimate}

\subsection{The \texorpdfstring{\zinvg}{Z (invisible) + gamma} and \texorpdfstring{\wlng}{W (leptonic) + gamma} background}

\label{subsec:vgamma}

The estimated per-bin SR yields of the \zinvg and \wlng backgrounds are scaled by transfer factors from the observed yields in the dilepton and single lepton CRs, respectively. The transfer factors are computed from simulation. In addition, to account for correlations between \PZ and \PW boson production, we compute from simulation a ratio that normalizes the \wlng yield to the \zinvg yield. Thus, the free background parameters in the likelihood function are the \zinvg SR bin yields, which account for both \PV+\PGg processes. This approach allows cancellation of theoretical uncertainties common to \PW\PGg and \PZ\PGg production, such as PDF and scale uncertainties.

The simultaneous likelihood fit used for signal extraction relies on accurate predictions of the ratios between the dominant backgrounds in the SR and CRs, as well as on the absolute normalization and shape of the \ETg distributions for the subdominant backgrounds. To achieve the most accurate possible predictions for these quantities, multiple correction factors are applied to each MC simulated sample of \PV+\PGg processes. The first correction uses the selection efficiency factor $\rho$~\cite{CMS:2015myp}, which compensates for minor variations between simulated and observed reconstruction and identification efficiencies across different particle candidates. This $\rho$ factor typically deviates from unity by only a few percent. A second adjustment implements higher-order QCD corrections, aligning the generator-level \ETg distributions with NNLO QCD calculations from the \DYRes program~\cite{Catani:2015vma}. Finally, a third modification incorporates NLO-EW effects (as taken from Ref.~\cite{Denner:2014bna,Denner:2015fca}) on \ETg distributions, derived from these references and updated with the LUXqed17 PDF set~\cite{Manohar:2017eqh}.

\subsection{Electron misidentification background}
\label{subsec:efake}

Another key background  originates from $\PW\to\Pe\PGn$ events where an electron is mistakenly identified as a photon. Misidentification occurs when hits in the upstream tracker layers along the electron trajectory are missed.
A seeding efficiency of $\epsilon$ for electrons with $\PT > 160\GeV$ is measured in data using a tag-and-probe~\cite{CMS:2011aa} technique in $\PZ \to \Pep\Pem$ events, and is validated through MC simulation. Misidentified electron events are modeled using a proxy sample of electron events, selected from the data by requiring an ECAL cluster with a pixel seed. These proxy events must satisfy the same selection criteria as those used for the signal candidate events except for the substitution of an electron for a photon candidate. 
The number of electron proxy events is then scaled by  $R_{\epsilon} = (1-\epsilon)/\epsilon$ to estimate the contribution from electron misidentification to our signal candidate selection. The ratio $R_{\epsilon}$ was measured to be $0.030\pm0.003$ and was found to be uniform across the considered \ETg spectrum. The dominant uncertainty in this estimate arises from the statistical uncertainty in the measurement of $\epsilon$.

\subsection{Jet misidentification background}
\label{subsec:jfake}

EM showers originating from hadronic activity can be misidentified as a photon signature. To estimate this background, the number of events are compared in two different subsets of a low--\ptmiss multijet data sample, where low--\ptmiss is defined as $\ptmiss < 30\GeV$. The ``first subset'' includes events with a photon candidate that meets the signal selection criteria, containing both true photons and jets that are misidentified as photons. The ``second subset'' consists of events where the  photon candidate satisfies less stringent shower shape requirements ($\sigetaeta > 0.0103$) and inverted isolation criteria compared to the signal candidates. The jet misidentification rate is then defined as the ratio of misidentified events in the first subset to the total number of events in the second subset.

The numerator is estimated by fitting the observed \sigetaeta distribution of the photon candidate in the first subset using a combination of simulated distributions and those derived from the observed data. For genuine photons, the shower width distribution is created using simulated $\PGg + \text{jets}$ events.
For jets that are misidentified as photons, the distribution is obtained from a  sample selected by inverting the charged-hadron isolation criteria and completely removing the shower shape requirement as both variables are correlated with each other.

The jet misidentification rate is measured to be between 0.01 to 0.16 across the \ETg spectrum, with relative uncertainties between 1.7 and 18.8\%. The systematic uncertainties in this measurement are dominant, comprising the shower shape distribution fit and shower shape modeling uncertainty, variations in the charged-hadron isolation threshold, low--\ptmiss requirement, and template bin width.

To estimate the background from jets misidentified as photons in the signal sample, the hadron misidentification rate is multiplied by the number of events in the high--\ptmiss CR, defined as $\ptmiss>200\GeV$. These events include a photon candidate that passes the same selection criteria as used for the second subset of the low--\ptmiss CR.

\subsection{Beam halo background}
\label{subsec:halo}
Beam halo background estimates are obtained by fitting the angular distributions of the calorimeter clusters. Energy clusters in the ECAL caused by beam halo muons are observed to concentrate around $\abs{\sin(\phi)} \sim 0$, whereas photon candidates from all other collision-related processes are uniformly distributed around $\phi$~\cite{Beamhalo}. This motivates the split of the SR into two parts according to $\phi$ to constrain the beam halo normalization as mentioned in Section~\ref{sec:event_selection}.

The partition of the SR can be thought of as a two-bin fit. Collision processes occupy the relative fractions of phase space in the horizontal ($H$) and  vertical ($V$) SR, $C_{H} = 1/\pi$ and $C_{V} = (\pi-1)/\pi$, respectively. For beam halo events, the corresponding fractions are determined by selecting a halo-enriched sample where the halo--veto identification criteria are inverted. Thus, a fit of the two subsets of the SR provides an estimate of the overall normalization of the beam halo background.

\subsection{Other minor SM background processes}
The SM \ttg, VV\PGg, \zllg, and \gj processes together account for about 10\% of the background processes in the SR. Although \zllg and \gj do not involve high-\pt invisible particles, the former can exhibit large \ptmiss when the leptons fail to be reconstructed, and the latter when jet energy is severely mismeasured. These backgrounds are estimated using MC simulation.

\section{Systematic uncertainties}
\label{sec:systematic}
The systematic uncertainty sources taken into account for the \ETg distribution ratios among the $\PV+\PGg$ processes are PDFs, higher-order QCD corrections, higher-order EW corrections, and data-to-simulation efficiency correction factors $\rho$. The PDF uncertainty is determined by varying the weight of each event using the weights provided in the NNPDF set, and taking the variance among the resulting \ETg distributions. This uncertainty is treated as fully correlated across different bins of \pt in the ratio between the \zinvg and \wlng processes, \ie, the variation of the ratio is constrained by the ratios of the upward and downward variations. Uncertainties related to higher-order QCD corrections are considered uncorrelated in the ratio between the \zinvg and \wlng processes. Since EW corrections become more significant at higher \ETg but are only known to NLO accuracy, their uncertainties are estimated using a particular method described in Ref.~\cite{Lindert:2017olm}, by assigning separate nuisance parameters to account for the uncertainty in the overall scale of the correction and its residual shape dependence with \ETg. Additionally, the full correction due to photon-induced $\PZ+\PGg$ and $\PW+\PGg$ production cross sections is considered as an uncertainty. Finally, the data-to-simulation correction factors $\rho$ for lepton identification efficiencies come with associated uncertainties that do not cancel out when calculating ratios between regions defined by different lepton selection criteria. All four uncertainties are treated as correlated across the \ETg bins.

Other smaller experimental uncertainties that affect the final background estimates include the integrated luminosity, jet and $\PGg$ energy scales, \ptmiss resolution, and data-to-simulation scale factors for object identification. Table~\ref{tab:systematics1} presents an overview of the systematic uncertainties considered in this analysis. In this table, ``Shape'' indicates uncertainties that impact the shape of the distributions. This means that the uncertainty can vary differently among bins in a histogram. Uncertainties affecting only the overall normalization (``Flat'') are modeled by nuisance parameters with a log-normal prior. A 3 to 11\% uncertainty is applied to account for the photon identification efficiency differences between data and simulation. The electron misidentification rate uncertainty (5--8\%) and jet misidentification rate (1--10\%) uncertaintiy both impact the shape of the estimated background distribution. Beam halo effects contribute to the halo background uncertainty (2--30\%), which is treated as a shape uncertainty due to its nonuniform distribution in the detector. This uncertainty is obtained by varying the selection used to construct the beam-halo control sample. Shape uncertainties are also considered for the jet energy scale (4--20\%) and photon energy scale (5.5--20\%). Lepton identification scale factor uncertainties (muon 1\%, electron 2\%) are treated as flat across both W and Z transfer factors, impacting only the normalization of the estimated background.  The integrated luminosity uncertainty is treated as a flat uncertainty, uniformly affecting all processes. It is estimated to be 1.2--2.5\%, partially correlated between the data-taking years~\cite{CMS:2021xjt,CMS:LUM-17-004,CMS:LUM-18-002}. Finally, MC and CR statistical uncertainties are treated bin-by-bin to account for statistical variations in the background estimation.

Among the listed uncertainties, theoretical uncertainties and those arising from photon and lepton identification efficiencies have the most significant impact on the signal extraction fit.
In the combined likelihood for the different data-taking years, systematic uncertainties are classified as either correlated or uncorrelated. Correlated uncertainties include jet energy scale, photon energy scale, and integrated luminosity, while uncorrelated uncertainties include factors such as photon identification and electron and jet misidentification rates, as well as theoretical uncertainties in the transfer factors.

\begin{table*}[!ht]
\centering
\topcaption{Summary of systematic uncertainties considered in the analysis. `\PW/\PZ Ratio' refers to correlated uncertainties propagated from \wlng to \zinvg background estimation (due to higher-order corrections and PDFs), `TF' denotes transfer factors, `MisID' represents misidentification backgrounds (electrons or jets misidentified as photons), and `Other' includes additional minor backgrounds.}
\label{tab:systematics1}
\cmsTable{
    \begin{scotch}{lccccccc}
        \shortstack[c]{Systematic \\ Uncertainty} &
        \shortstack[c]{\PW/\PZ \\ Ratio} &
        \shortstack[c]{\PW \\ TF} &
        \shortstack[c]{\PZ \\ TF} &
        \shortstack[c]{Other \\ Background} &
        \shortstack[c]{Electron \\ MisID} &
        \shortstack[c]{Jet \\ MisID} &
        Halo \\[\cmsTabSkip]
\hline
        NNLO QCD, EW, PDF & \shortstack[c]{Shape \ (1--10\%)} &  \NA &  \NA &  \NA &  \NA &  \NA &  \NA \\[\cmsTabSkip]
        Photon ID SF &  \NA &  \NA &  \NA & \shortstack[c]{Shape \\ (3--11\%)} &  \NA &  \NA &  \NA \\[\cmsTabSkip]
        MisID rate &  \NA &  \NA &  \NA &  \NA & \shortstack[c]{Shape \\ (5--8\%)} & \shortstack[c]{Shape \\ (1--10\%)} &  \NA \\[\cmsTabSkip]
        Halo &  \NA &  \NA &  \NA &  \NA &  \NA &  \NA & \shortstack[c]{Shape \\ (2--30\%)} \\[\cmsTabSkip]
        Jet energy scale &  \NA &  \NA &  \NA & \shortstack[c]{Shape \\ (4--20\%)} &  \NA &  \NA &  \NA \\[\cmsTabSkip]
        Photon energy scale &  \NA &  \NA &  \NA & \shortstack[c]{Shape \\ (5.5--20\%)} &  \NA &  \NA &  \NA \\[\cmsTabSkip]
        Integrated luminosity &  \NA & \shortstack[c]{Flat \\ (1--2\%)} & \shortstack[c]{Flat \\ (1--2\%)} & \shortstack[c]{Flat \\ (1--2\%)} &  \NA &  \NA &  \NA \\[\cmsTabSkip]
        Lepton identification & \shortstack[c]{Flat \\ (1--2\%)} & \shortstack[c]{Flat \\ (1--2\%)} & \shortstack[c]{Flat \\ (1--2\%)} &  \NA &  \NA &  \NA &  \NA \\[\cmsTabSkip]
        MC and CR statistical &  \NA &  \NA &  \NA & \shortstack[c]{Bin-by-bin} &  \NA &  \NA & \NA
    \end{scotch}
}
\end{table*}

\section{Results}
\label{sec:results}

\subsection{Signal extraction}
\label{sec:extraction}

The potential signal contribution is extracted from the data by performing a simultaneous fit to the \ETg distributions in the SR and CRs. Uncertainties in several quantities are modeled as nuisance parameters in the fitting process. 

The free parameter of the likelihood  function is the signal strength modifier (\Pgm), which, for a given signal hypothesis, controls the signal normalization relative to the model cross section.

\subsection{Post-fit distributions}

\begin{figure*}[th]
  \centering

     \includegraphics[width=0.48\textwidth]{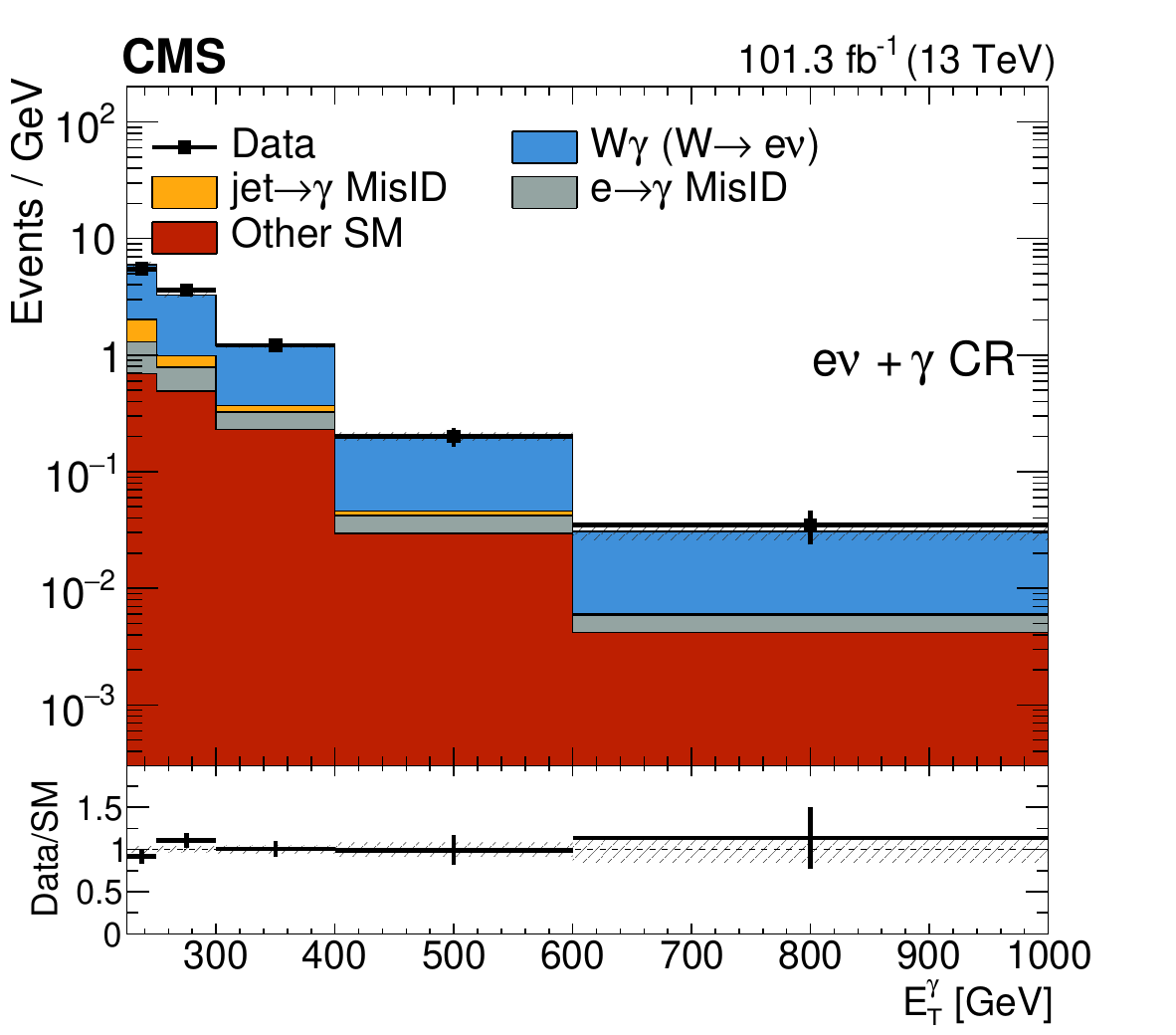}
     \includegraphics[width=0.48\textwidth]{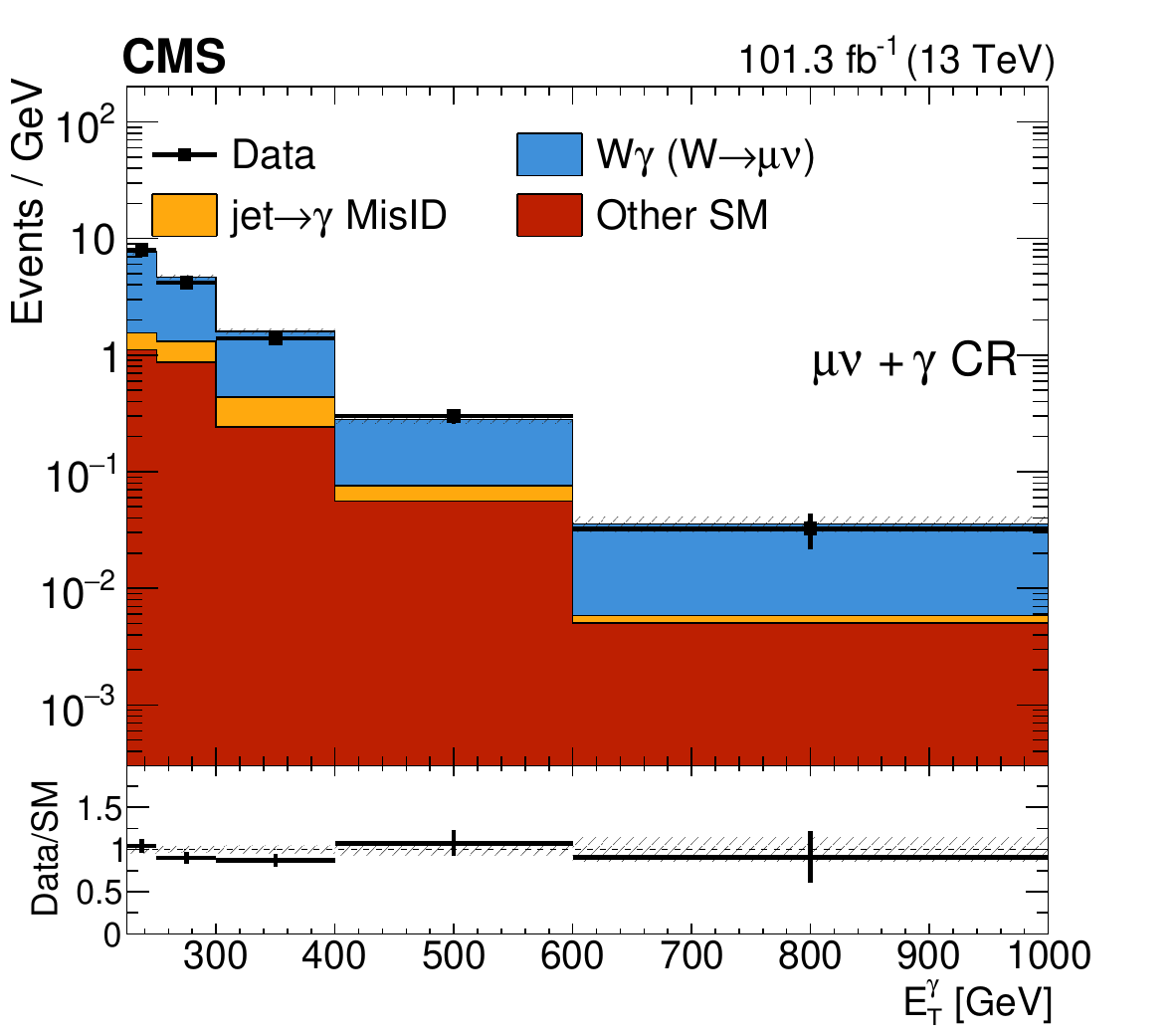}
    
    \caption{
      Comparison of data and background post-fit distributions in the $\Pe \nu + \PGg$ (left) and $\Pgm \nu +\PGg$ (right) CRs for the combination of 2017 and 2018 data sets. The last bin of the distribution includes the overflow events. The ratios of data to the background predictions are shown in the lower panels, with the cross hatched uncertainty bands including the combination of all systematic uncertainties.
      }
    \label{fig:CR1}
\end{figure*}

\begin{figure*}[th]
  \centering

    \includegraphics[width=0.48\textwidth]{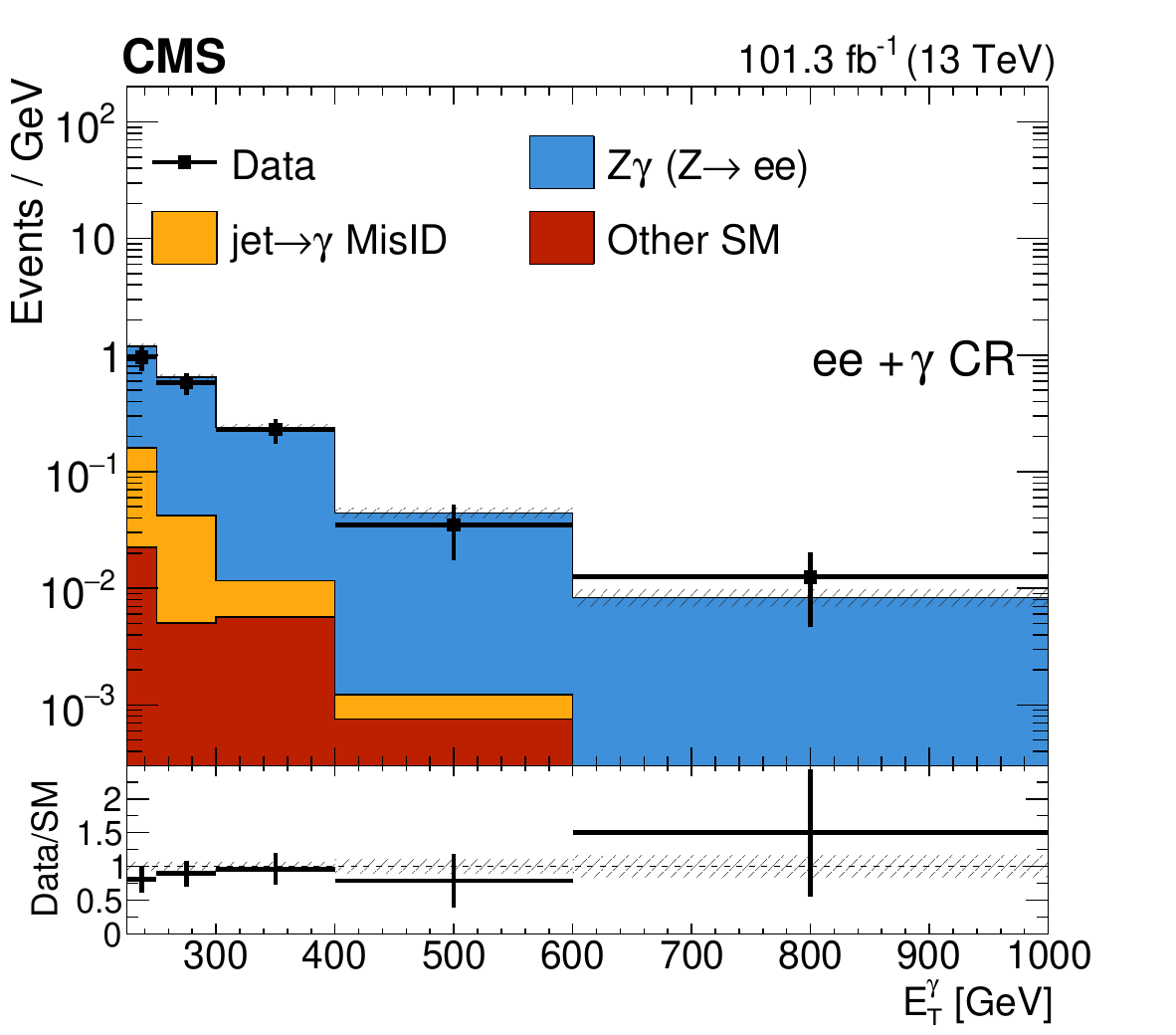}
    \includegraphics[width=0.48\textwidth]{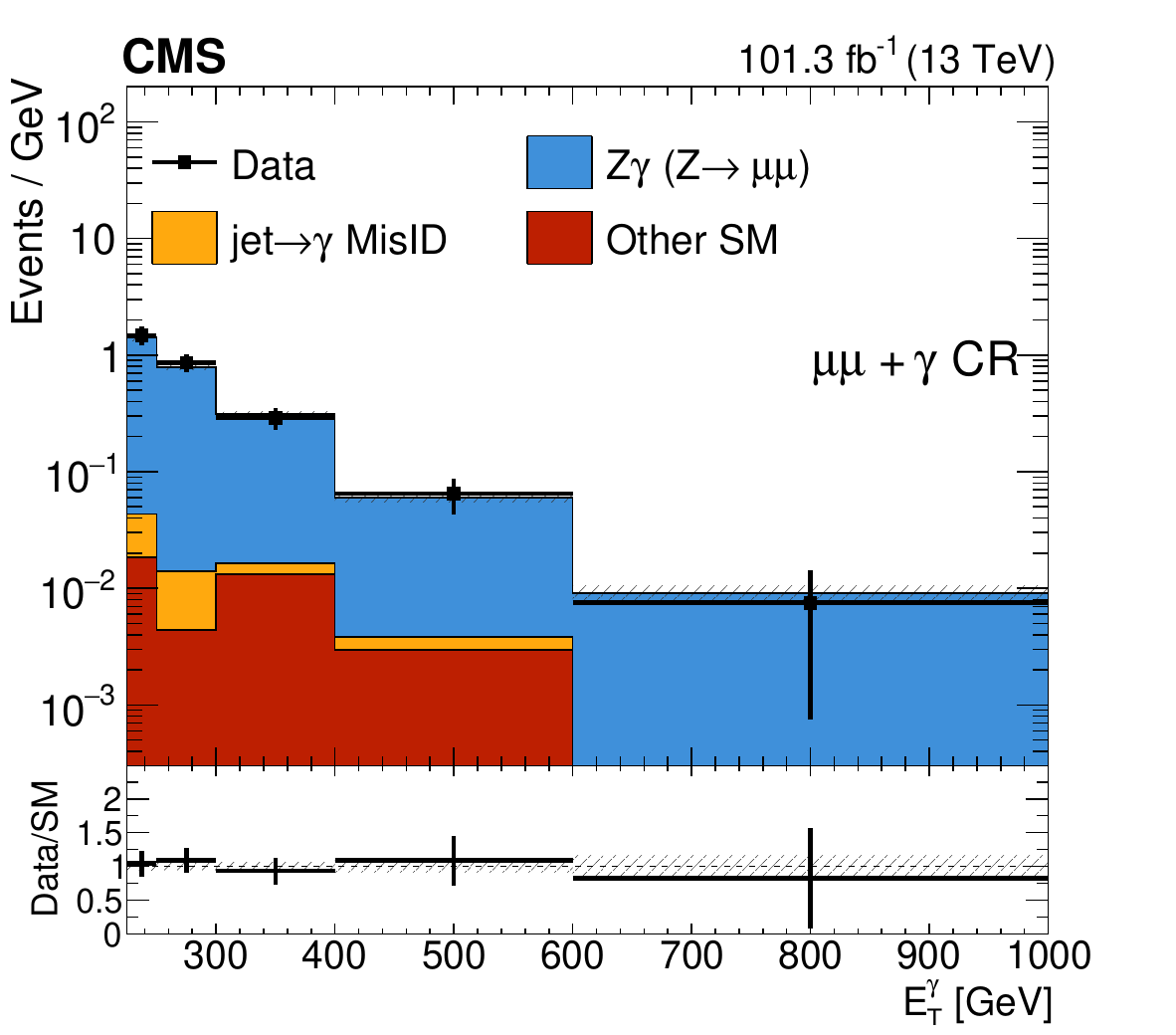}
    \caption{
    Comparison of data and background post-fit distributions in the $\Pe\Pe + \PGg$ (left) and $\Pgm\Pgm+ \PGg$ (right) CRs for the combination of 2017 and 2018 data sets. The last bin of the distribution includes the overflow events. The ratios of data to the background predictions are shown in the lower panels, with the cross hatched uncertainty bands including the combination of all systematic uncertainties.
    }   
    \label{fig:CR2}
\end{figure*}

\begin{figure*}[th]
  \centering
    \includegraphics[width=0.48\textwidth]{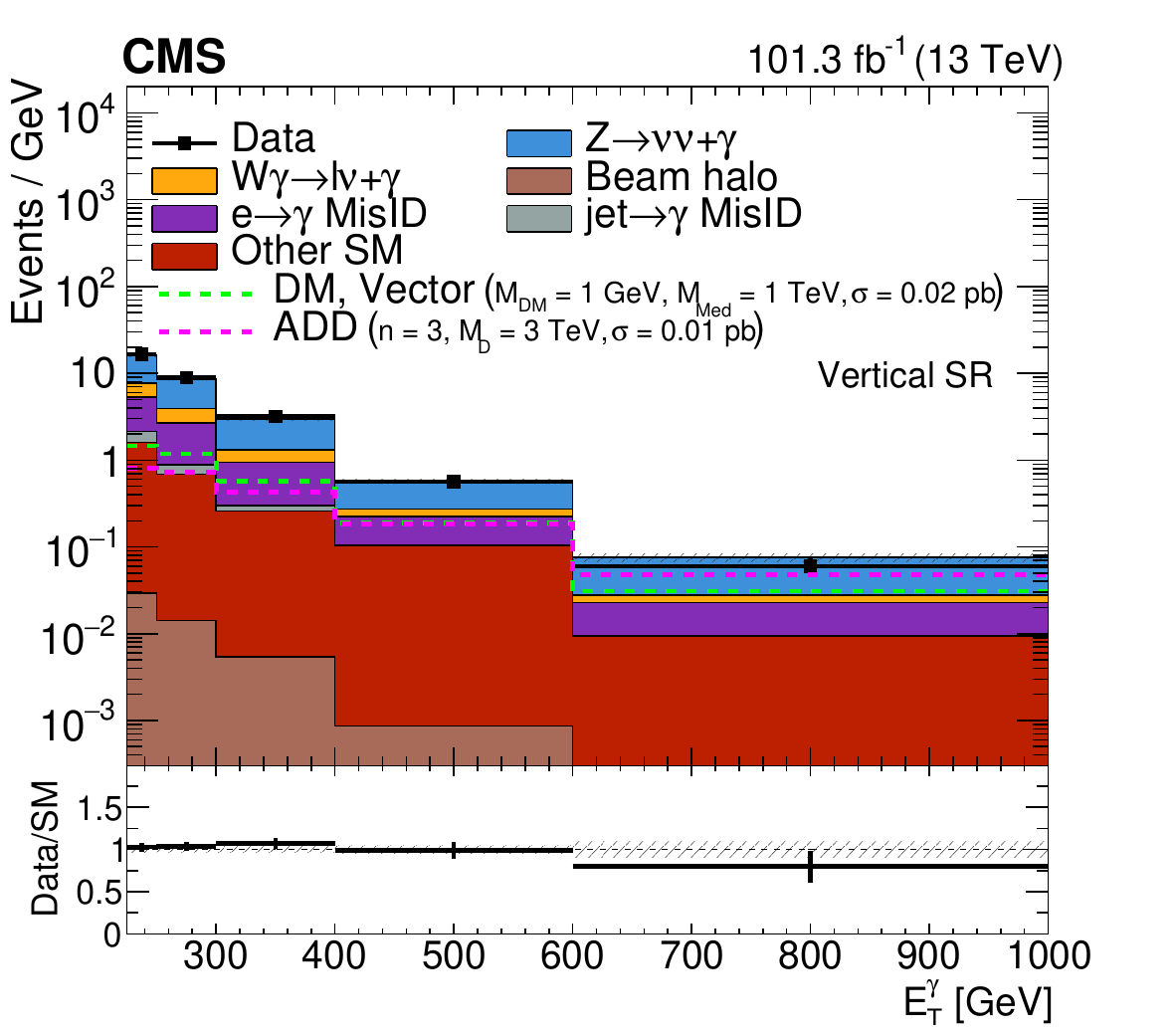}
     \includegraphics[width=0.48\textwidth]{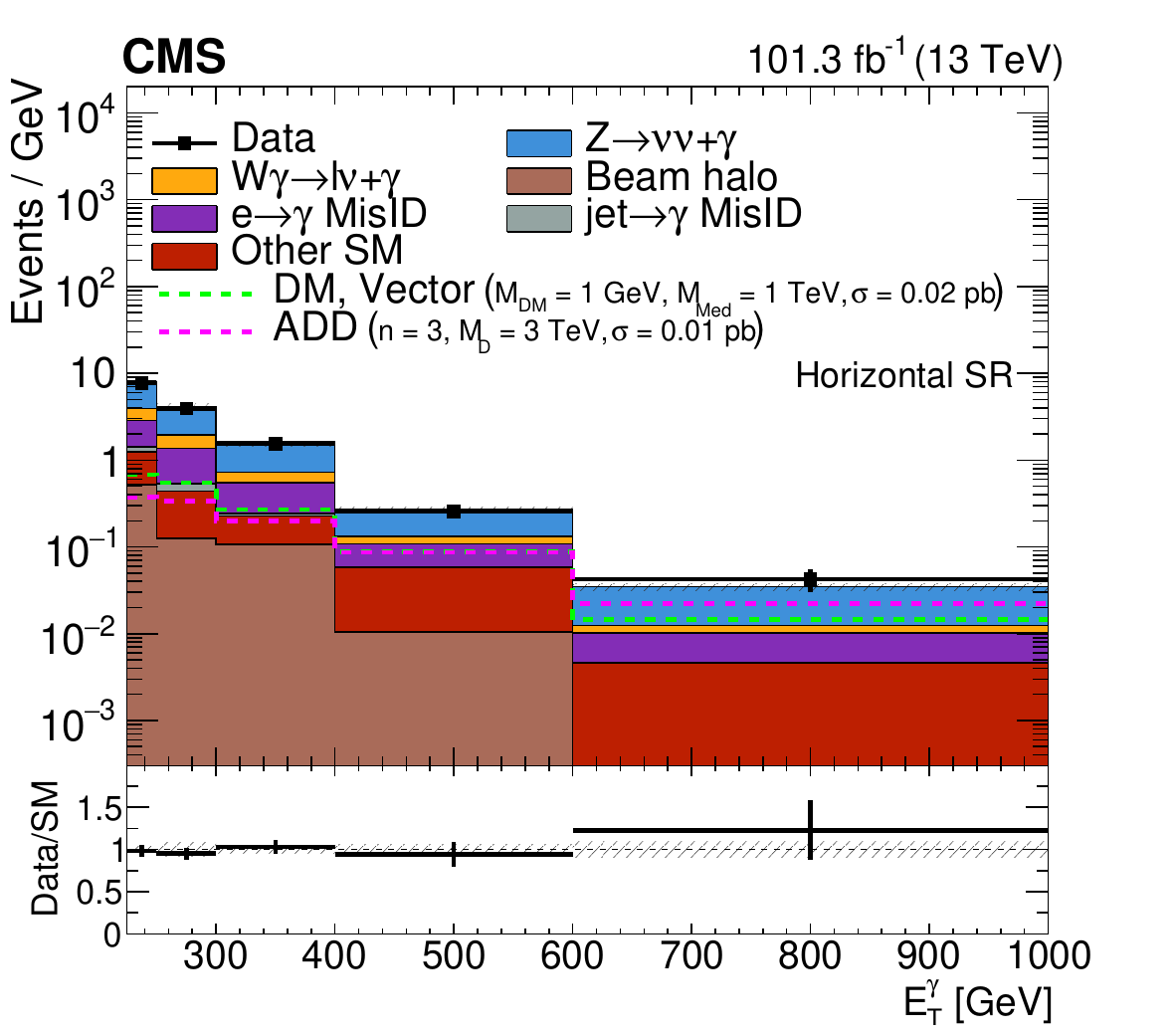}

\caption{
    Comparison of data and background post-fit distributions for the vertical (left) and  horizontal (right) regions for the combination of 2017 and 2018 data sets. Templates for signal hypotheses are shown overlaid as light green and magenta dashed lines along with their cross section values. The last bin of the distribution includes the overflow events.
    }
    \label{fig:SR1}
\end{figure*}

The result is extracted using the maximum likelihood fit approach provided by the CMS statistical analysis tool \textsc{Combine}~\cite{CMS:2024onh}. Figures~\ref{fig:CR1} and~\ref{fig:CR2} show the observed post-fit distributions of \ETg in the four CRs, while Fig.~\ref{fig:SR1} shows the expected \ETg distributions in the vertical and horizontal SR using the combined 2017 and 2018 data set. Table~\ref{tab:BkGSummary} shows the corresponding total yields, categorized by process, in the vertical and horizontal SR.

\begin{table*}[!ht]
\newcolumntype{x}{D{,}{\,\pm\,}{4.8}}
\newcolumntype{y}{D{,}{\,\pm\,}{3.2}}
\centering
\normalsize
\topcaption{Total yield in vertical and horizontal regions using the combined 2017 and 2018 data set.}
\label{tab:BkGSummary}

\begin{scotch}{l x y x y}
Process
& \multicolumn{2}{c}{Pre-fit}
& \multicolumn{2}{c}{Post-fit} \\[\cmsTabSkip]

& \multicolumn{1}{c}{Vertical}
& \multicolumn{1}{c}{Horizontal}
& \multicolumn{1}{c}{Vertical}
& \multicolumn{1}{c}{Horizontal} \\[\cmsTabSkip]
\hline

\zinvg & 621,33 & 290,16 & 688,20 & 321,9 \\
\wlng  & 174,7  & 81,3   & 172,5  & 80,3 \\
Electron misidentification & 257,7 & 117,3 & 262,7 & 120,3 \\
QCD    & 28,1.0 & 11,0.2 & 28,1   & 11,0.2 \\
Beam halo & 1.85,0 & 26,21 & 2.2,0.2 & 32,18 \\
\ttg/\tg & 28,2 & 13,1 & 28,2 & 13,1 \\
Other background & 96,5 & 45,2 & 94,5 & 43,2 \\[\cmsTabSkip]

Total background & 1204,35 & 583,27 & 1275,22 & 622,21 \\[\cmsTabSkip]

ADD(\mD = 3, n = 3)   & 155 & 72 & 0 & 0 \\
DM(\mmed = 1, \mdm = 1)  & 203 & 95 & 0 & 0 \\[\cmsTabSkip]

Data & 1315 & 613 & 1315 & 613 \\

\end{scotch}
\end{table*}

The final result is obtained by performing a simultaneous fit that combines the previously published 2016 analysis~\cite{Sirunyan:2017ewk} with the 2017 and 2018 data sets. These data sets are fitted using a common likelihood function, while the systematic uncertainties, including those carried over from the 2016 analysis, are detailed in Section~\ref{sec:systematic}.

\subsection{Limits and constraints}
No significant excess of events beyond the SM expectation is observed. Upper limits are determined for the production cross section of three new physics processes mentioned in Section~\ref{sec:introduction}. For each model, a 95\% confidence level (\CL) upper limit is obtained utilizing the asymptotic \CLs criterion~\cite{Junk:1999kv,Read:2002hq,Cowan:2010js}, with a test statistic based on the negative logarithm of the likelihood in Section~\ref{sec:extraction}.

\begin{figure*}[th]
  \centering
   \includegraphics[width=0.48\textwidth]{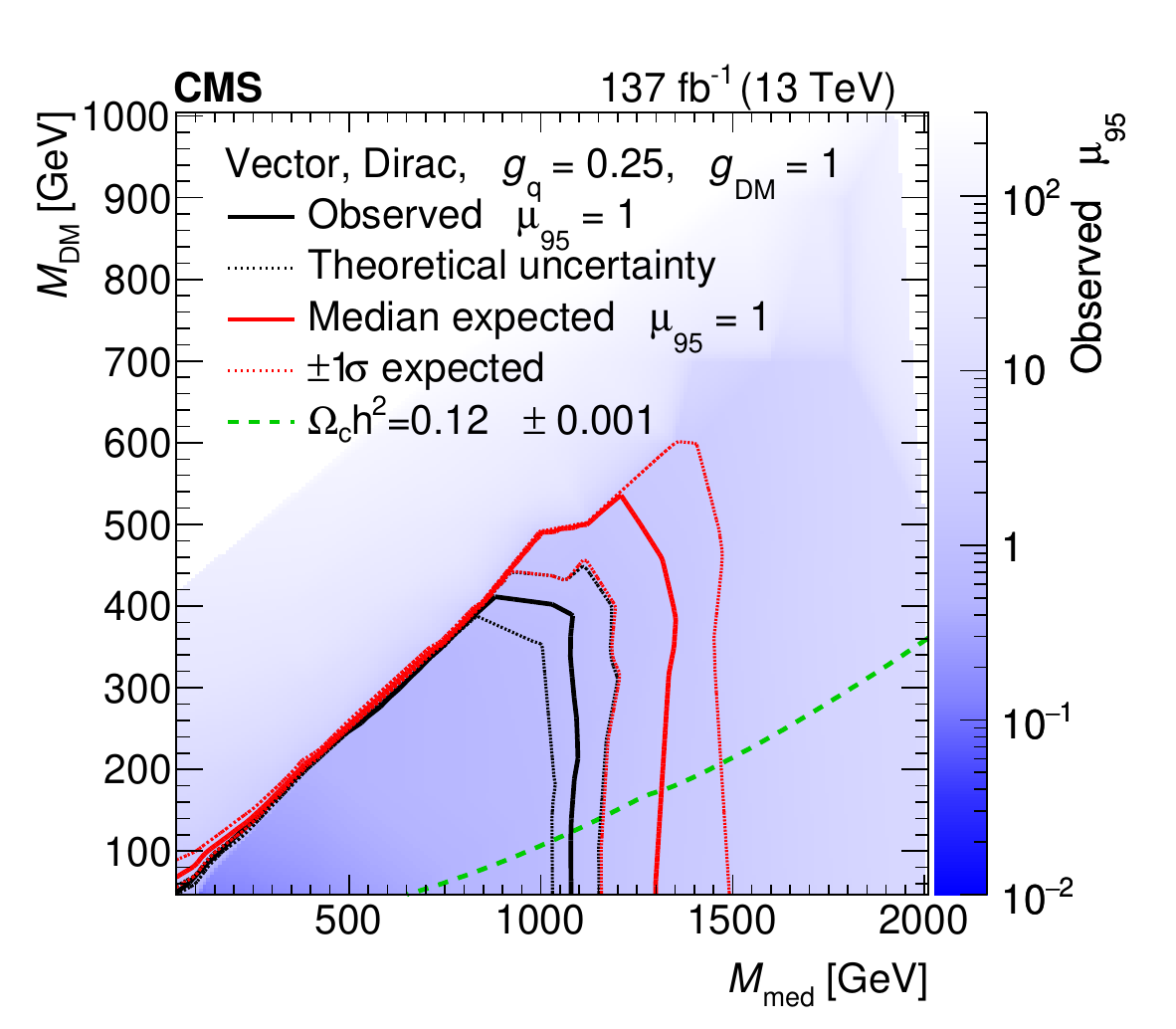}
   \includegraphics[width=0.48\textwidth]{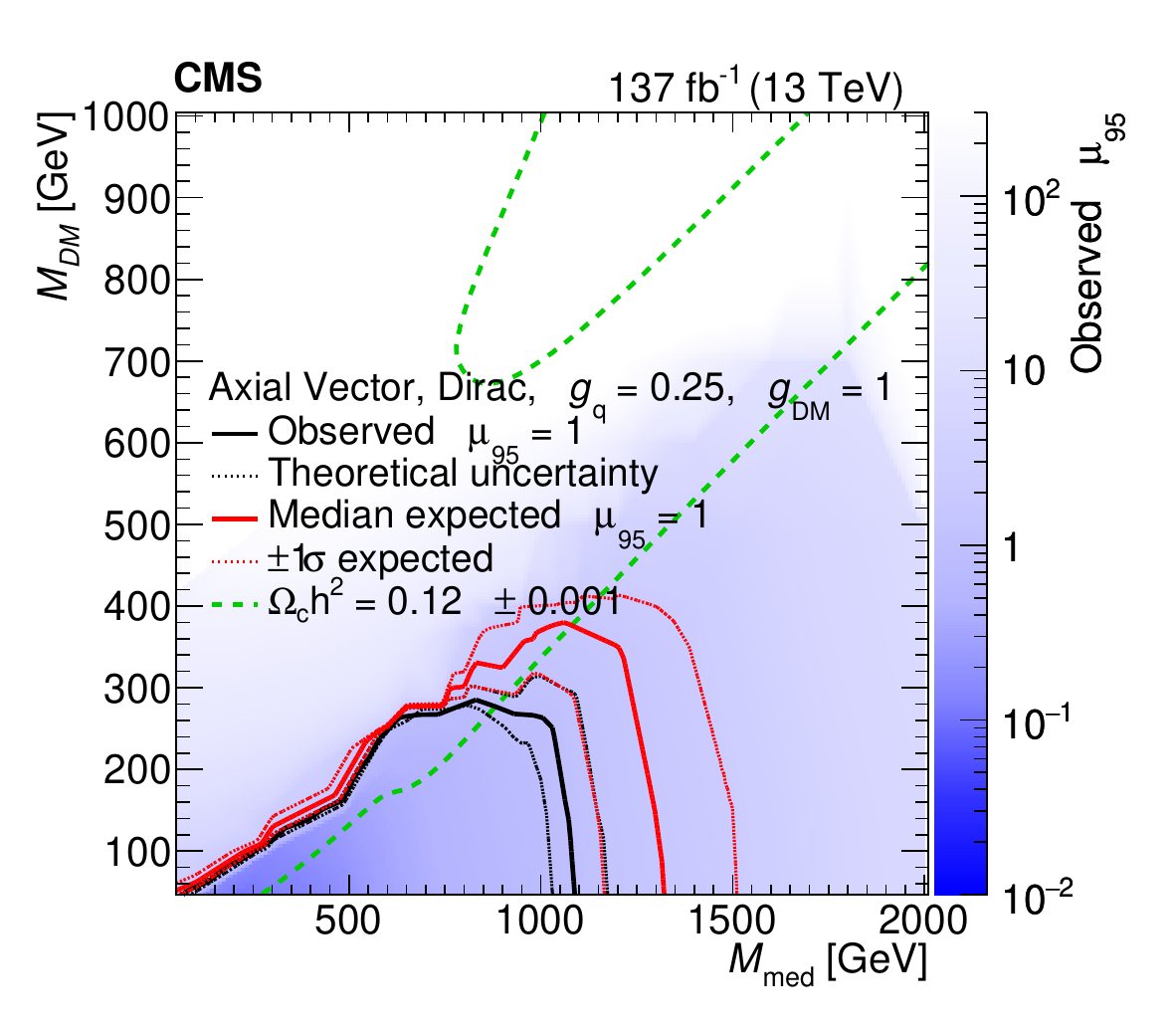}
   
    \caption{
    The ratio of 95\% \CL upper cross section limits to the theoretical cross section ($\mu_{95}$), for simplified DM models with vector (left) and axial-vector (right) mediators, using the full 2016--2018 data set corresponding to an integrated luminosity of \currentL\fbinv, assuming \gq = 0.25 and \gDM = 1. Expected $\mu_{95} = 1$ contours are overlaid in red. The region below and to the left of the observed contour is excluded. For simplified DM model parameters in the region below the lower green dash contour, and also above the corresponding upper contour in the right hand plot, cosmological DM abundance exceeds the density observed by the Planck satellite experiment. }
    \label{fig:2d}
\end{figure*}

The simplified DM model parameters  proposed by the ATLAS-CMS Dark Matter Forum Ref.~\cite{Abercrombie:2015wmb} are intended to enable the comparison and translation of various DM search results. Figure~\ref{fig:2d} illustrates the 95\% \CL upper cross section limits relative to the corresponding theoretical cross section ($\mu_{95}= \sigma_{95\%}/\sigma_{\text{theory}}$) for the  vector and axial-vector mediator scenarios in the \mmed--\mdm plane. The solid red (black) curves represent the expected (observed) contour where $\mu_{95} = 1$. The dotted line above (or below) the expected contour indicates the $\pm 1\sigma$ uncertainty band on the expected limit. For the simplified DM NLO models considered, mediator masses up to 1085\GeV are excluded for \mdm values less than 50\GeV for both vector and axial vector mediator. The apparent difference between the good agreement observed in Fig.~\ref{fig:SR1} and the observed limits shown in Fig.~\ref{fig:2d} arises because Fig.~\ref{fig:SR1} presents the 2017--2018 post-fit SR distributions, whereas Fig.~\ref{fig:2d} is based on the combined 2016--2018 data sets. The combined observed limit reflects statistical fluctuations present in the 2016 data-taking period.

Results for vector and axial-vector mediators are then compared with constraints established by the cosmic microwave background measurements of the cosmological relic density of DM by the Planck satellite experiment~\cite{Ade:2015xua}. For each model separately, the expected DM abundance is estimated using the thermal freeze-out mechanism implemented in the {\sc MadDM} framework~\cite{Backovic:2013dpa} and is then compared to the observed cold DM density, $\Omega_ch^{2} = 0.12\pm0.001$~\cite{Ade:2015xua}, where $\Omega_c$ represents the DM relic abundance and $h$ is the dimensionless Hubble constant.

\begin{figure*}[th]
  \centering
    \includegraphics[width=0.48\textwidth]{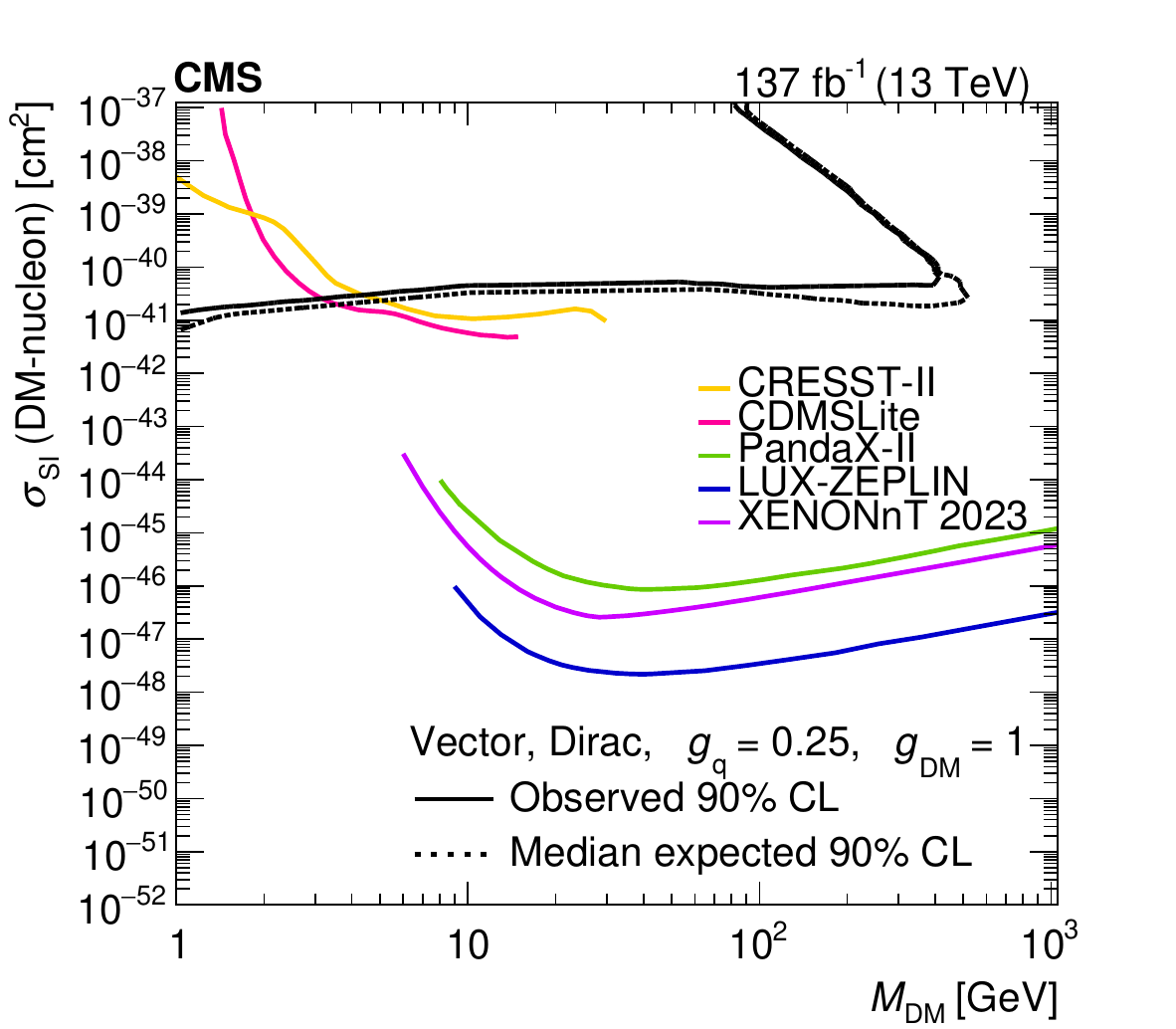}
    \includegraphics[width=0.48\textwidth]{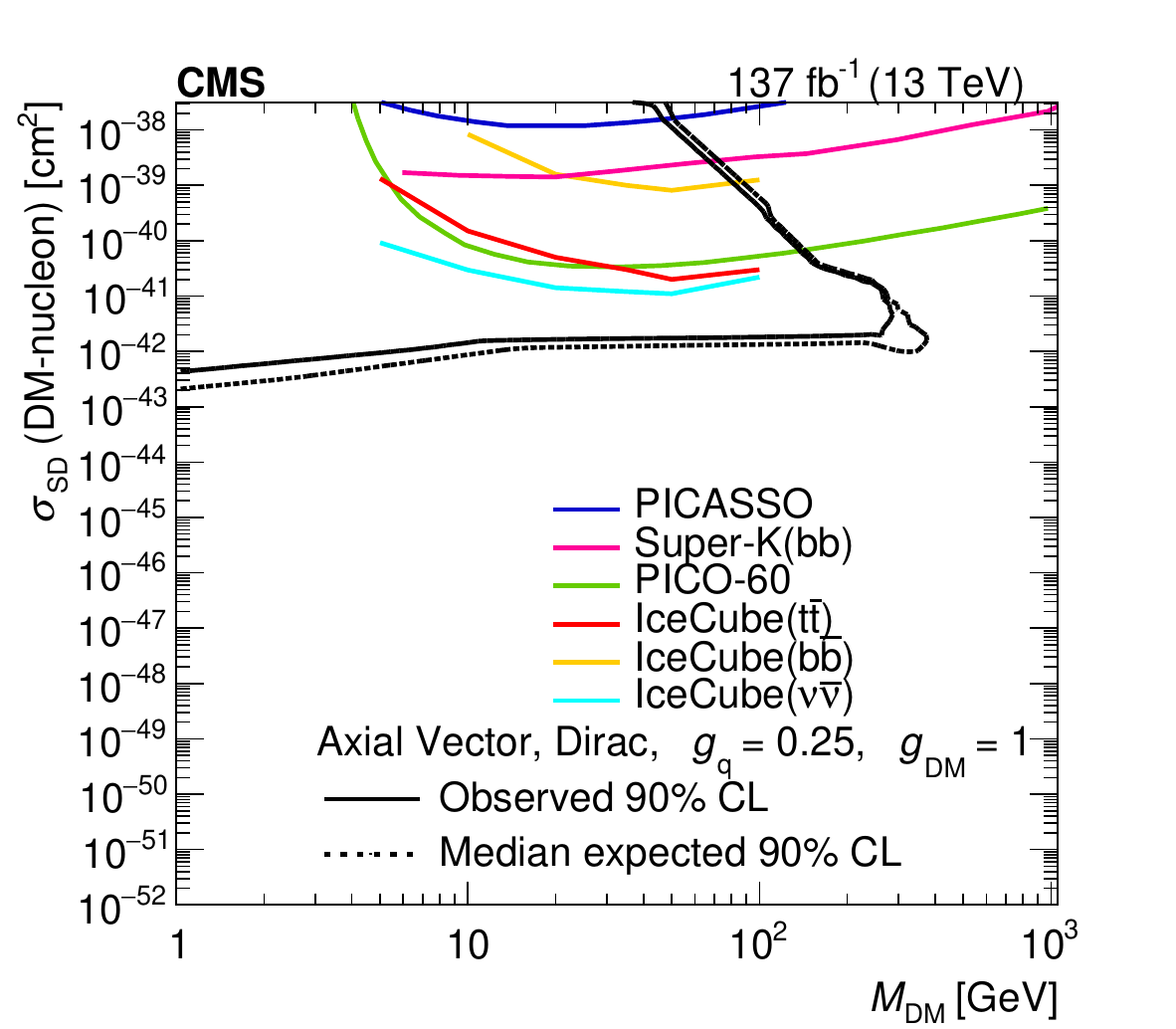}
    \caption{ The 90\% \CL exclusion limits on the \PGc--nucleon spin-independent (left) and spin-dependent (right) scattering cross sections involving vector and axial-vector operators, respectively, are shown as a function of \mdm, using the 2016--2018 data set. Simplified model DM parameters \gq = 0.25 and \gDM = 1 are assumed. Also shown are corresponding exclusion contours, where regions above the curves are excluded, from recent results by the CDMSLite~\cite{Agnese:2015nto}, LUX-ZEPLIN~\cite{LZ:2024zvo}, PandaX-II~\cite{PandaX-II:2017hlx}, XENONnT~\cite{XENON:2023cxc}, CRESST-II~\cite{Angloher:2015ewa}, PICO-60~\cite{PICO:2017tgi}, IceCube~\cite{IceCube:2021xzo}, PICASSO~\cite{Behnke:2016lsk}, and Super-Kamiokande~\cite{Choi:2015ara} Collaborations.}
    \label{fig:cont}
\end{figure*}

The exclusion contours in Fig.~\ref{fig:2d} are converted into the $\sigma_{\text{SI}}, \sigma_{\text{SD}}$--\mdm plane, where $\sigma_{\text{SI}}$ and $\sigma_{\text{SD}}$ represent the spin-independent and spin-dependent DM--nucleon scattering cross sections, respectively, as depicted in Fig.~\ref{fig:cont}. The procedure for this translation and the presentation of results follows the prescription outlined in Ref.~\cite{Boveia:2016mrp}. Specifically, to facilitate a direct comparison with direct-detection experiments, these limits are computed at 90\% \CL ~\cite{Abercrombie:2015wmb}. Compared to direct-detection experiments, this search sets more stringent limits for DM masses under 2\GeV in the spin-independent scenario and below 200\GeV in the spin-dependent case.

\begin{figure}[th]
\centering

 \includegraphics[width=0.5\textwidth]{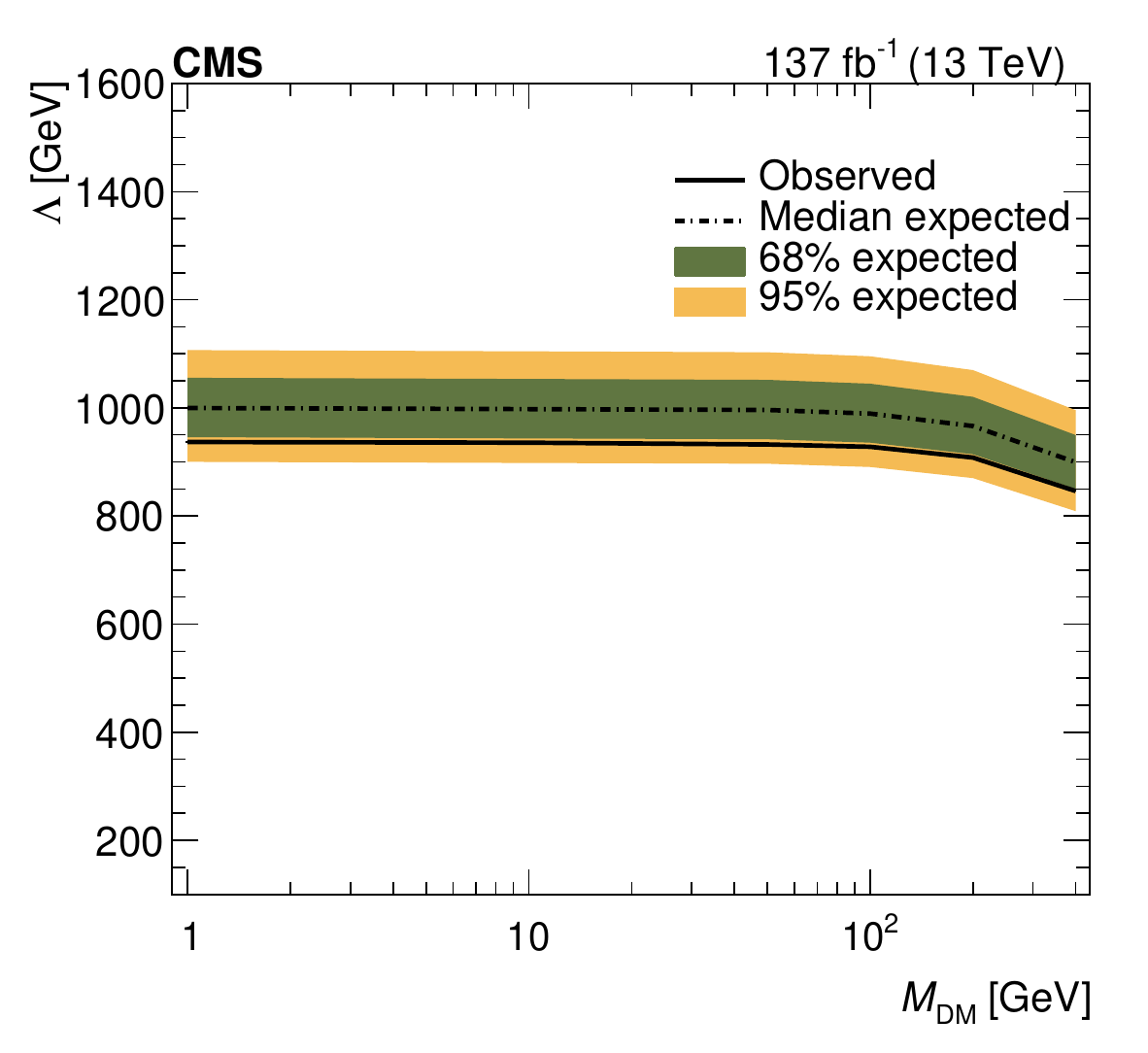}
\caption{The 95\% \CL observed and expected lower limits on $\Lambda$ for an effective EW--DM contact interaction, as a function of \mdm using the 2016--2018 data set corresponding to an integrated luminosity of \currentL\fbinv.}\label{fig:DMEWKlimits}
\end{figure}

For the DM model characterized by a contact interaction of type $\PGg\PGg\PGc\PAGc$, upper limits are established on the production cross section, which are subsequently translated into lower limits on the suppression scale $\Lambda$ for $k_{1} = k_{2} = 1.0$. The observed (expected) 95\% \CL lower limits on $\Lambda$ as a function of the DM mass \mdm are shown in Fig.~\ref{fig:DMEWKlimits}. For \mdm values ranging from 1 to 800\GeV, $\Lambda$ values up to 940\GeV are excluded at 95\% \CL, with an expected exclusion up to 1000\GeV, using the data set corresponding to an integrated luminosity of \currentL\fbinv. 

\begin{figure}[th]
  \centering
    \includegraphics[width=0.5\textwidth]{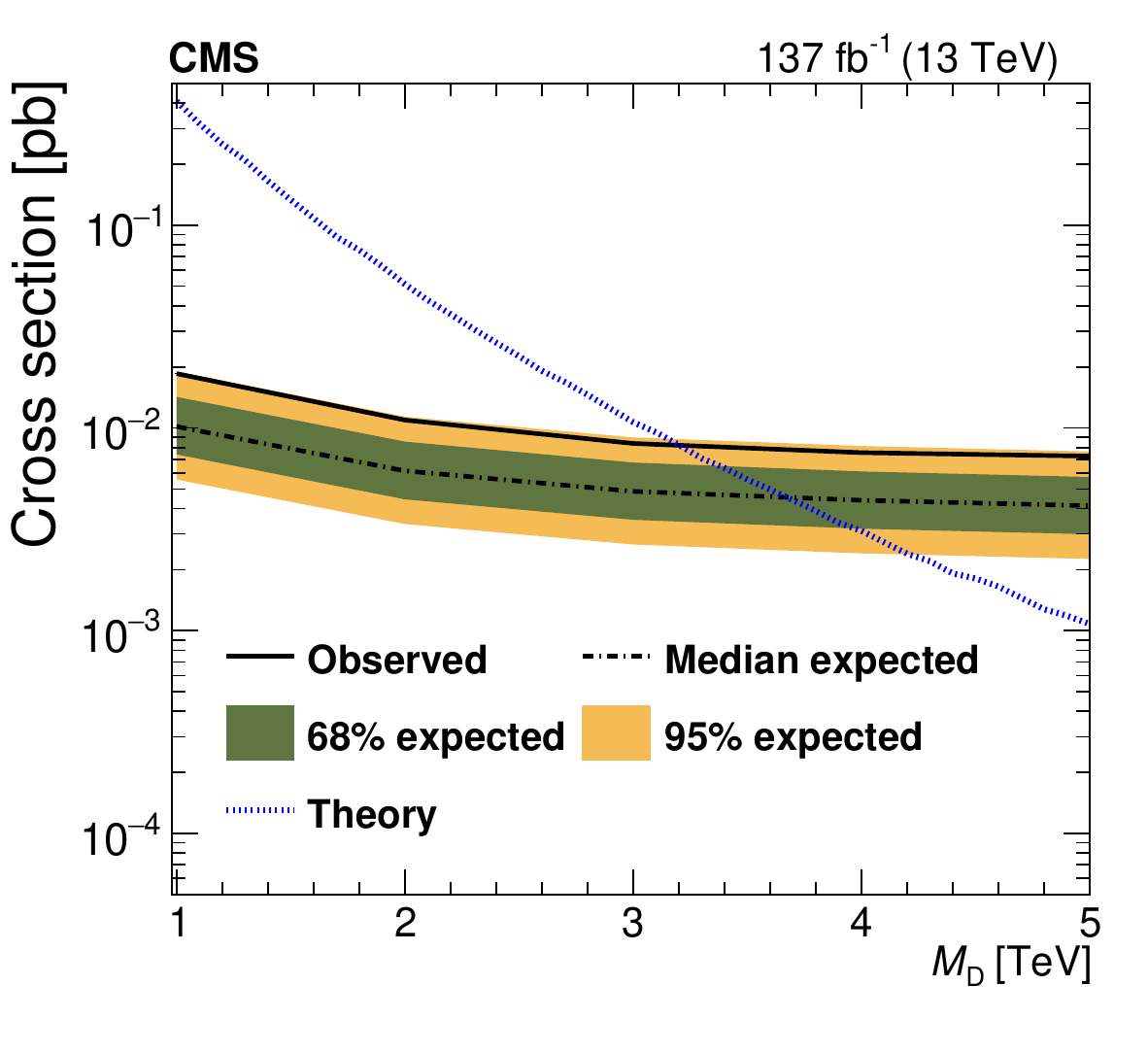}
    \caption{The 95\% \CL upper limits on the ADD graviton production cross section as a function of \mD, for $n=3$ extra dimensions.}
    \label{fig:ADDLimits}
\end{figure}

\begin{figure}[th]
  \centering

   \includegraphics[width=0.5\textwidth]{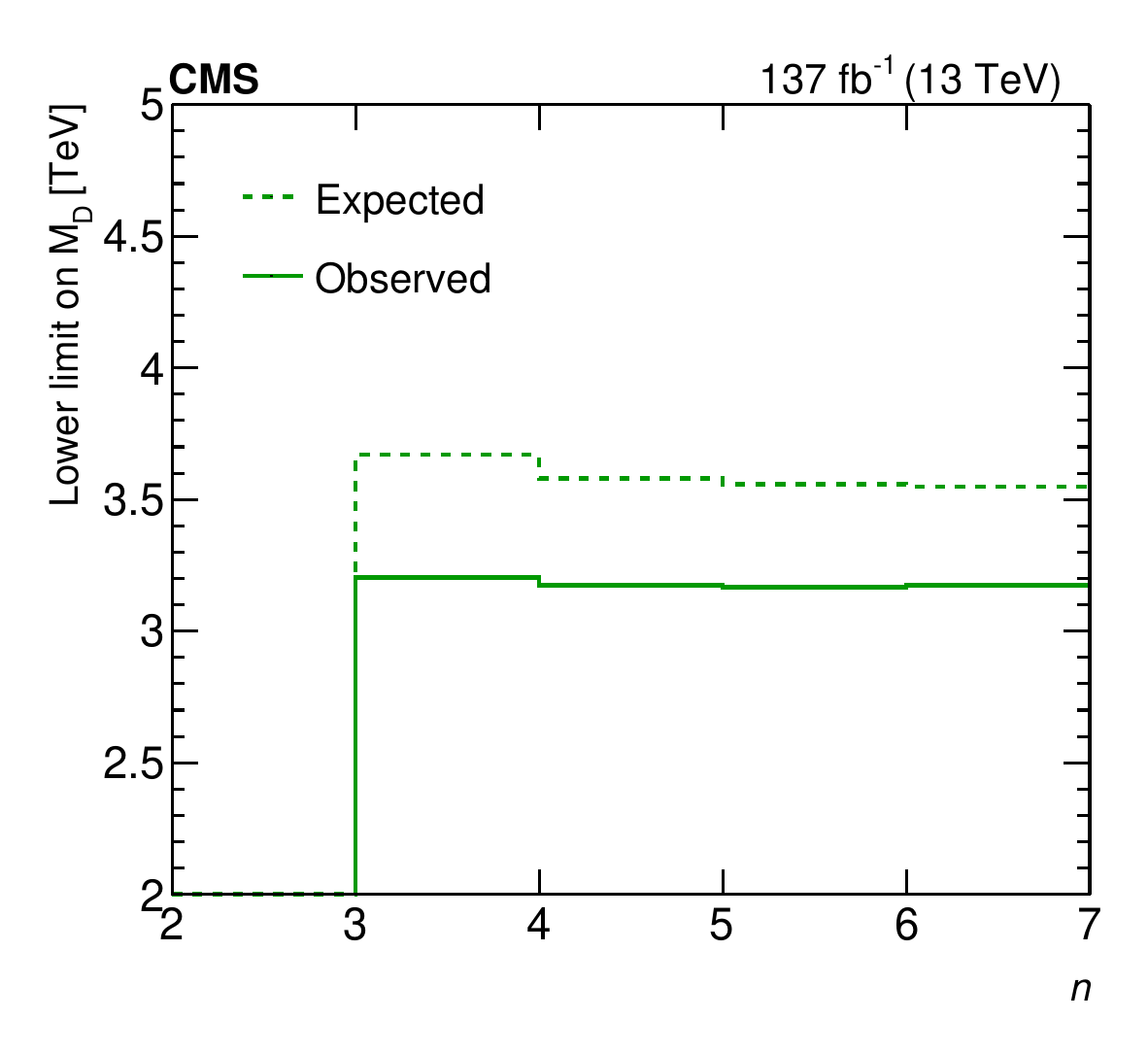}
    \caption{Lower limit on the fundamental Planck scale \mD as a function of the number of extra dimensions $n$, using the 2016-2018 data sets with an integrated luminosity of \currentL\fbinv.}
    \label{fig:MDLimits}
\end{figure}

Figure~\ref{fig:ADDLimits} shows the upper limit and the theoretically calculated ADD graviton production cross section for $n$ = 3 extra dimensions, as a function of \mD . The lower limit on \mD as a function of $n$ extra dimensions is shown in Fig.~\ref{fig:MDLimits}. The limit is nearly constant across extra dimensions ($n$ = 3 to $n$ = 6), excluding \mD values up to 3.2 (3.7)\TeV, observed (expected), with the current analysis, providing stronger constraints on large extra dimensions than previous searches. 

These limits provide important constraints on simplified DM models, effective EW--DM contact interactions, and large extra dimension scenarios using the monophoton final state at the LHC.

\section{Summary}

\label{sec:summary}

Proton-proton collisions producing a high transverse momentum photon and large missing transverse momentum have been investigated to search for new phenomena, using a data set corresponding to \currentL\fbinv of integrated luminosity recorded at $\sqrt{s} = 13\TeV$ at the LHC. No deviations from the standard model predictions are observed. For the simplified dark matter production models considered, the observed (expected) lower limit on the mediator mass is 1085 (1300)\GeV in both cases for a 1\GeV dark matter particle mass. For an effective electroweak dark matter contact interaction, the observed (expected) lower limit on the suppression parameter $\Lambda$ is 940 (1000)\GeV, which is an improvement of 10\% (5\%) over the 2016 analysis, while for the model with extra spatial dimensions, values of the effective Planck scale \mD up to 3.2\TeV are excluded between 3 and 6 extra dimensions, improving upon the 2016 results by 10\% (11\%). These results set the most stringent limits to date on these parameters of the electroweak contact interaction and extra dimension models in the monophoton final state. The improvement in the expected limits obtained with the full 2016--2018 data set on the parameters of the model is modest, of the order of 10--14\% (30--40\% in the expected cross sections limits) compared to 2016. This occurs for two reasons. Firstly, the increased instantaneous luminosity of the collider in the later data-taking years necessitated higher thresholds on \ptmiss and \ETg in the trigger. The corresponding adjustments in the event selection result in a reduction of acceptance, and of the lever arm for distinguishing signal from background in the shape of the \ETg distributions. Secondly, because systematic uncertainties, mainly the theoretical QCD scale and electroweak uncertainties in the \zinvg and \wlng backgrounds, become more significant relative to the diminished statistical uncertainties of the full sample.

\begin{acknowledgments}
We congratulate our colleagues in the CERN accelerator departments for the excellent performance of the LHC and thank the technical and administrative staffs at CERN and at other CMS institutes for their contributions to the success of the CMS effort. In addition, we gratefully acknowledge the computing centers and personnel of the Worldwide LHC Computing Grid and other centers for delivering so effectively the computing infrastructure essential to our analyses. Finally, we acknowledge the enduring support for the construction and operation of the LHC, the CMS detector, and the supporting computing infrastructure provided by the following funding agencies: SC (Armenia), BMBWF and FWF (Austria); FNRS and FWO (Belgium); CNPq, CAPES, FAPERJ, FAPERGS, and FAPESP (Brazil); MES and BNSF (Bulgaria); CERN; CAS, MoST, and NSFC (China); MINCIENCIAS (Colombia); MSES and CSF (Croatia); RIF (Cyprus); SENESCYT (Ecuador); ERC PRG, TARISTU24-TK10 and MoER TK202 (Estonia); Academy of Finland, MEC, and HIP (Finland); CEA and CNRS/IN2P3 (France); SRNSF (Georgia); BMFTR, DFG, and HGF (Germany); GSRI (Greece); NKFIH (Hungary); DAE and DST (India); IPM (Iran); SFI (Ireland); INFN (Italy); MSIT and NRF (Republic of Korea); MES (Latvia); LMTLT (Lithuania); MOE and UM (Malaysia); BUAP, CINVESTAV, CONACYT, LNS, SEP, and UASLP-FAI (Mexico); MOS (Montenegro); MBIE (New Zealand); PAEC (Pakistan); MES, NSC, and NAWA (Poland); FCT (Portugal);  MESTD (Serbia); MICIU/AEI and PCTI (Spain); MOSTR (Sri Lanka); Swiss Funding Agencies (Switzerland); MST (Taipei); MHESI (Thailand); TUBITAK and TENMAK (T\"{u}rkiye); NASU (Ukraine); STFC (United Kingdom); DOE and NSF (USA).

\hyphenation{Rachada-pisek} Individuals have received support from the Marie-Curie program and the European Research Council and Horizon 2020 Grant, contract Nos. 675440, 724704, 752730, 758316, 765710, 824093, 101115353, 101002207, 101001205, and COST Action CA16108 (European Union); the Leventis Foundation; the Alfred P. Sloan Foundation; the Alexander von Humboldt Foundation; the Science Committee, project no. 22rl-037 (Armenia); the Fonds pour la Formation \`a la Recherche dans l'Industrie et dans l'Agriculture (FRIA) and Fonds voor Wetenschappelijk Onderzoek contract No. 1228724N (Belgium); the Beijing Municipal Science \& Technology Commission, No. Z191100007219010, the Fundamental Research Funds for the Central Universities, the Ministry of Science and Technology of China under Grant No. 2023YFA1605804, the Natural Science Foundation of China under Grant No. 12061141002, 12535004, and USTC Research Funds of the Double First-Class Initiative No. YD2030002017 (China); the Ministry of Education, Youth and Sports (MEYS) of the Czech Republic; the Shota Rustaveli National Science Foundation, grant FR-22-985 (Georgia); the Deutsche Forschungsgemeinschaft (DFG), among others, under Germany's Excellence Strategy -- EXC 2121 ``Quantum Universe" -- 390833306, and under project number 400140256 - GRK2497; the Hellenic Foundation for Research and Innovation (HFRI), Project Number 2288 (Greece); the Hungarian Academy of Sciences, the New National Excellence Program - \'UNKP, the NKFIH research grants K 131991, K 133046, K 138136, K 143460, K 143477, K 146913, K 146914, K 147048, 2020-2.2.1-ED-2021-00181, TKP2021-NKTA-64, and 2021-4.1.2-NEMZ\_KI-2024-00036 (Hungary); the Council of Science and Industrial Research, India; ICSC -- National Research Center for High Performance Computing, Big Data and Quantum Computing, FAIR -- Future Artificial Intelligence Research, and CUP I53D23001070006 (Mission 4 Component 1), funded by the NextGenerationEU program (Italy); the Latvian Council of Science; the Ministry of Education and Science, project no. 2022/WK/14, and the National Science Center, contracts Opus 2021/41/B/ST2/01369, 2021/43/B/ST2/01552, 2023/49/B/ST2/03273, and the NAWA contract BPN/PPO/2021/1/00011 (Poland); the Funda\c{c}\~ao para a Ci\^encia e a Tecnologia, grant CEECIND/01334/2018 (Portugal); the National Priorities Research Program by Qatar National Research Fund;  MICIU/AEI/10.13039/501100011033, ERDF/EU, "European Union NextGenerationEU/PRTR", and Programa Severo Ochoa del Principado de Asturias (Spain); the Chulalongkorn Academic into Its 2nd Century Project Advancement Project, the National Science, Research and Innovation Fund program IND\_FF\_68\_369\_2300\_097, and the Program Management Unit for Human Resources \& Institutional Development, Research and Innovation, grant B39G680009 (Thailand); the Kavli Foundation; the Nvidia Corporation; the SuperMicro Corporation; the Welch Foundation, contract C-1845; and the Weston Havens Foundation (USA).
\end{acknowledgments}

\bibliography{auto_generated}

\cleardoublepage \appendix\section{The CMS Collaboration \label{app:collab}}\begin{sloppypar}\hyphenpenalty=5000\widowpenalty=500\clubpenalty=5000\cmsinstitute{Yerevan Physics Institute, Yerevan, Armenia}
{\tolerance=6000
A.~Hayrapetyan, V.~Makarenko\cmsorcid{0000-0002-8406-8605}, A.~Tumasyan\cmsAuthorMark{1}\cmsorcid{0009-0000-0684-6742}
\par}
\cmsinstitute{Institut f\"{u}r Hochenergiephysik, Vienna, Austria}
{\tolerance=6000
W.~Adam\cmsorcid{0000-0001-9099-4341}, J.W.~Andrejkovic, L.~Benato\cmsorcid{0000-0001-5135-7489}, T.~Bergauer\cmsorcid{0000-0002-5786-0293}, M.~Dragicevic\cmsorcid{0000-0003-1967-6783}, C.~Giordano, P.S.~Hussain\cmsorcid{0000-0002-4825-5278}, M.~Jeitler\cmsAuthorMark{2}\cmsorcid{0000-0002-5141-9560}, N.~Krammer\cmsorcid{0000-0002-0548-0985}, A.~Li\cmsorcid{0000-0002-4547-116X}, D.~Liko\cmsorcid{0000-0002-3380-473X}, M.~Matthewman, I.~Mikulec\cmsorcid{0000-0003-0385-2746}, J.~Schieck\cmsAuthorMark{2}\cmsorcid{0000-0002-1058-8093}, R.~Sch\"{o}fbeck\cmsAuthorMark{2}\cmsorcid{0000-0002-2332-8784}, D.~Schwarz\cmsorcid{0000-0002-3821-7331}, M.~Shooshtari\cmsorcid{0009-0004-8882-4887}, M.~Sonawane\cmsorcid{0000-0003-0510-7010}, W.~Waltenberger\cmsorcid{0000-0002-6215-7228}, C.-E.~Wulz\cmsAuthorMark{2}\cmsorcid{0000-0001-9226-5812}
% ...
\par}
\vskip\cmsinstskip
\dag:~Deceased\\
$^{1}$Also at Yerevan State University, Yerevan, Armenia\\
$^{2}$Also at TU Wien, Vienna, Austria\\
% ...
\end{sloppypar}
%%% END EDITABLE REGION %%%
% skeleton_end
\end{document}